\documentclass[11pt,preprint]{aastex}
\usepackage{epsfig}
\usepackage{ulem}
\usepackage{color}
\usepackage{textcomp}

\def\oiii{{[O \sc iii]}\,}

\def\nii{{[N \sc ii]}\,}

\def\dvr{\Delta V_{\rm r}}
\def\dvb{\Delta V_{\rm b}}

\def\ppagn{$pp$-AGNs}
\def\ppgalaxies{$pp$-galaxies}

\def\kms{$\rm km~s^{-1}$}
\def\ergs{$\rm erg~s^{-1}$}

\def\pp{\prime\prime}

\begin{document}
\title{Double-peaked narrow emission-line galaxies from Sloan Digital Sky Survey. \\
I. Sample and basic properties}
\author{
Jun-Qiang Ge\altaffilmark{1},
Chen Hu\altaffilmark{1},
Jian-Min Wang\altaffilmark{1,2,*},
Jin-Ming Bai\altaffilmark{3}
and Shu Zhang\altaffilmark{1}
}

\altaffiltext{1}{
Key Laboratory for Particle Astrophysics, Institute of High Energy Physics,
Chinese Academy of Sciences, 19B Yuquan Road, Beijing 100049, China
}
\altaffiltext{2}{
National Astronomical Observatories, Chinese Academy of Sciences, 20 Datun Road, Beijing 100020,
China}

\altaffiltext{3}{Yunnan Observatory, Chinese Academy of Sciences, Kunming 650011, China}
\altaffiltext{*}{Corresponding author: wangjm@mail.ihep.ac.cn}

\begin{abstract}
Recently, much attention has been given to double-peaked narrow emission-line galaxies, some of which are
suggested to be related with merging galaxies.  We make a systematic
search to build the largest sample of these sources from Data Release 7 of Sloan Digital Sky Survey (SDSS
DR7). With reasonable criteria of fluxes, full-width-half-maximum of emission lines and separations of the
peaks, we select 3,030 double-peaked narrow emission-lines galaxies. In light of the existence of broad Balmer
lines and the locations of the two components of double-peaked narrow emission lines distinguished by the
Kauffmann et al. (2003) criteria in the Baldwin-Phillips-Terlevich (BPT) diagram, we find that there are
81 type I AGN, 837 double-type II AGN (2-type II), 708 galaxies with double star forming components (2-SF),
400 with mixed star forming and type II AGN components (type II + SF) and 1,004 unknown-type objects. As
a by-product, a sample of galaxies (12,582) with asymmetric or top-flat profiles of emission
lines is established. After inspecting the SDSS images of the two samples visually, we find 54 galaxies with dual cores.
The present samples can be used to study the dynamics of merging galaxies, the triggering mechanism of black
hole activity, the hierarchical growth of galaxies and dynamics of narrow line regions driven by outflows
and rotating disk.
\end{abstract}

\section{Introduction}
Double-peaked narrow emission-line active galactic nuclei (hereafter \ppagn) have been found since the
1980s (e.g. Heckman et al. 1981, 1984; Keel 1985; Whittle 1985a,b,c), but were mainly interpreted as
an indicator of outflows from active nuclei or rotating narrow line regions (e.g. Greene \& Ho 2005).
Since the suggestion that double-peaked profiles could be produced by dual AGN (Zhou et al. 2004; Wang
et al. 2009), much attention has been recently paid to
search for individual objects and study statistical properties of \ppagn\ samples in light of several
fundamental issues related with evolution of galaxies and AGN (for a summary of previous works on this
subject see Table 1). There are 8 individual objects
with double-peaked narrow emission-line profiles in total listed in Table 1, and three samples that
made systematic searches for \ppagn\ from SDSS DR7 (Wang et al. 2009; Liu et al. 2010a; Smith et al. 2010;
hereafter W09, L10 and S10, respectively). W09, L10 and S10 select their samples by employing
different selection criteria, including restricting equivalent width (EW),
redshift, signal-to-noise ratio (S/N), narrow emission line (NEL) flux ratio etc.
W09 and L10 select 190 \ppagn\ in total from SDSS galaxy sample using Kewley et al. (2001) criterion
in the BPT diagram (Baldwin et al. 1981). This paper applies new selection scheme and criteria to hunt
double-peaked narrow emission-line galaxies (\ppgalaxies) from $\sim 920,000$ galaxies.

Double-peaked narrow line profiles can be caused by different mechanisms discussed subsequently. First, bi-polar
outflows driven by starbursts or AGN radiation pressure have been suggested for many years, especially
in several individual objects, such as NGC 1068 (Crenshaw \& Kraemer 2000; Das et al. 2006, 2007),
NGC 4151 (Hunchings et al. 1998; Crenshaw et al. 2000; Das et al. 2005), Mrk 3 (Ruiz et al. 2001;
Crenshaw et al. 2010a), Mrk 78 (Fisher et al. 2011), and Mrk 573 (Schlesinger et al. 2009; Fisher et
al. 2010). In particular, outflows can produce double-peaked (e.g. Mrk 78 see Fisher et al. 2011)
or multiple-peaked emission-line profiles (e.g. NGC 1068, see also Crenshaw et al. 2010b) depending on
the projection of the outflows in the line of sight. However, W09 found an anti-correlation between the
flux ($F_{\rm r}/F_{\rm b}$) and shift ($\Delta \lambda_{\rm r}/\Delta \lambda_{\rm b}$) ratios of the
double peaks, where $F$ and $\Delta \lambda$ are the fluxes and shifts of red and blue components, which
has been confirmed by Liu et al. (2010a, see their Figure 6{\it e}), Smith et al. (2010, see their Figure 5)
and Fu et al. (2011, see their Figure 5). This new statistical relation strongly suggests co-rotating dual
AGN model. Unless the bi-polar outflows with very asymmetric mass rates are under momentum conservation,
the anti-correlation would not favor the presence of bi-polar outflows (W09)\footnote{Here we assume that
the outflows are optically thin. The radiation from them is proportional to the mass of the emitting gas
expected.}. Second, rotating disk-like narrow line regions (NLRs) with circular orbits are {\it not}
able to produce the anti-correlation because a large fraction of \ppagn\ have $F_{\rm r}/F_{\rm b}\neq 1$,
unless the rotating NLRs have complicated elliptical orbits (or inhomogeneous mass distribution over the
rotating disk) and orientations to observers. Third, the origin of double-peaked profiles could be
interpreted by dual AGN which will be in agreement with the anti-correlation (W09; L10; Smith et al. 2010;
Rosario et al. 2011) as identified by some works (e.g. Fu, et al. 2011; Shen, et al. 2011) listed in Table 1.
In order to identify the nature of these $pp$-AGN, near infrared (NIR) images and optical long slit
spectroscopy have shown that about 50\% type I $pp$-AGN (Fu et al. 2011), 10-20\% type II $pp$-AGN (Liu et
al. 2010b; Shen et al. 2011; Fu et al. 2011, 2012), or more than $50\%$ \ppagn\ (Comerford et al. 2011) and
50\% double-peaked quasars (Rosario et al. 2011) have two cores. The nature of the remaining objects is open
(a single core: bi-polar outflows or rotating NLR; or spatially unresolved dual-cores).

Double-peaked narrow emission-line galaxies play an important role in studies of the dynamics
of AGN NLRs and merging galaxies. The formation of NLRs is insufficiently understood even though they
are spatially resolved in some individual objects (Bennert et al. 2002; Schmitt et al. 2003; Bennert et al.
2006a,b). Jet-induced outflows are responsible for NLR formation in some AGN (Bicknell
et al. 1998) in light of individual objects, such as Mrk 3 (Capetti et al. 1999), and Mrk 78 (Whittle
\& Wilson 2004). More clear evidence for the jet-clouds interaction in NGC 4051 has been revealed by
{\it Chandra} (Wang et al. 2011). Cold clouds are formed by the cocoons produced through the Kelvin-Helmholtz
instability of the relativistic jet propagating in the interstellar medium (ISM) (Steffen et al. 1997). NLRs
driven by the intermediate scale of outflows or jets can cause the appearance of double-peaked profiles
(Rosario et al. 2008, 2010). This is supported by the fact that there is a
correlation between the jet power and width of NELs (Wilson \& Willis 1980) and powerful,
compact linear radio sources have anomalously broad \oiii\ lines usually (Whittle 1992). It remains to be
systematically researched on NLR dynamics driven by jet-induced outflows in a large and homogeneous sample.
Furthermore, feedback to starburst through AGN outflows is presumed to play a key role in regulating bulge
growth, but it still needs to be evidenced by observations of a large \ppgalaxies\ sample driven by bi-polar
outflows. On the other hand, Greene \& Ho (2005) conducted a
systematic investigation of NLR dynamics from a large SDSS sample. They draw a conclusion that the NLR
is mainly governed by the potential of bulges though there are some evidence for rotation-supported
NLR in light of Integral Field Unit (IFU) rotation curves (Vega Beltr\'an et al. 2001; Dumas et al. 2007; Stoklasov\'a
et al. 2009). Clearly, those \ppagn with double-peaked NELs produced by a rotation-supported NLR would be used to justify if
the rotation-supported NLR follows the dynamics of bulges (Dumas et al. 2007; Ho 2009) or the rotation
of partial disks of galaxies, even provide an opportunity to study the relative orientation of AGN and
host galaxies (Shen et al. 2010; Lagos et al. 2011).

Merging galaxies enter a special phase in which double cores exist after their galactic disks merge
within the framework of the theory of hierarchical galaxy formation and evolution (e.g. Longair 2008;
Yu et al. 2011). As illustrated by SDSS images (e.g. Figure 2 in Darg et al. 2010), there is a
sequence of galaxy mergers from tidal interaction between two galaxies
with a separation of tens of kpc to an ``approaching post-merger" stage containing
dual cores with a $\sim$kpc separation. If hierarchical growth of galaxies works,
are there AGN-starburst cores, or galaxies with dual starbursts? How often?
Do they appear as \ppgalaxies? If so, these \ppgalaxies\ could open a new clue to understanding the
issues related to triggering starbursts and black hole activity, and help to understand the details
of mergers. Second, dual-cored galaxies are important for the subject of black hole binaries
searched through shifts between broad lines and narrow lines (Tsalmantza et al. 2011, Eracleous et al. 2011)
since dual-cored galaxies are progenitors of binary black holes. With the dual-cored galaxies,
we will have an entire scenario of merging process. These motivate us to
systematically search for \ppgalaxies\ to look into statistical properties.

In this paper, we make an attempt to establish the largest sample of \ppgalaxies\ through a systematical
search from SDSS DR7 spectra. As a by-product, we also build a sample of asymmetric or
top-flat narrow line (asym-NEL) galaxies, some of which could be candidates of \ppgalaxies. Section 2 describes
the followed procedures, selection criteria and the final samples of this work. We systematically
inspect images of the \ppgalaxies\ and asym-NEL galaxies in Section 3 and find 54
of them with dual cores. In Section 4, we
present the basic properties of the sample. We draw conclusions in Section 5. We use a standard
cosmology with $H_0=71~{\rm km~s^{-1}~Mpc^{-1}}$, $\Omega_M=0.27$ and $\Omega_{\Lambda}=0.73$ in this paper.

\section{Sample Selection}
For the goal of selecting \ppgalaxies,
we develop a series of automatic algorithms in light of Monte-Carlo simulations given by the Appendix A.
We apply the algorithms to SDSS galaxy spectra and get candidates of the \ppgalaxies. Then we visually
inspect the candidate spectra to establish the
final sample. We follow the procedures: (1) select emission-line galaxies from the entire sample
of galaxies spectroscopically observed by SDSS; (2) pure emission-line spectra are created by subtracting
stellar components using spectral synthesis methods; (3) {\it F}-test is employed to examine whether
double-Gaussians are needed to fit each NEL in individual objects, if so, they are selected as
candidates of \ppgalaxies; (4) finally, we visually inspect spectra of the candidates and establish
the final sample. We classify the sample into sub-samples according to the width of their broad Balmer
emission lines and NEL ratios using the BPT diagram (Baldwin et al. 1981). Figure 1 shows a
brief flowchart of the whole procedures which are described in detail below.

\subsection{Emission-line spectra of galaxies}
We start from the database of the MPA-JHU\footnote{The MPA/JHU catalog can be downloaded from
http://www.mpa-garching.mpg.de/SDSS/DR7. We use H$\alpha$, ~~\oiii\, and H$\beta$ measurements of
the MPA sample.} SDSS DR7 galaxy sample (in total 927,552 galaxy spectra; Abazajian et al. 2009).
First, 737,824 sources are selected from those with a signal-to-noise ratio ${\rm S/N}>9$ based on
the results by L10\footnote{It should be noted that ${\rm S/N}>5$ is employed by
L10 as a selection criterion (see Table 1), but their resultant \ppagn\ actually
all have ${\rm S/N}>9$. Hence we use ${\rm S/N}>9$ in this paper.}
in the rest-frame wavelength range 4800-5100\AA. Second, 337,188 emission-line galaxies are
selected from the 737,824 sources in light of that at least one of  H$\alpha$,
\oiii $\lambda 5007$ and H$\beta$
lines has equivalent width greater than 3\AA \ (Hao et al. 2005).

Measurement accuracy of redshifts of galaxies is very important to investigate properties of
\ppgalaxies. In this work, we measure cosmological redshifts ($z$) of the galaxy absorption lines
(i.e. Ca {\sc ii} H+K and Mg {\sc i}b triplets) using the
direct pixel-fitting method following Greene \& Ho (2006) and Ho et al. (2009). The stellar
templates used here are selected from the Indo-U.S. Library of Coud\'{e} Feed Stellar Spectra
(Valdes et al. 2004, hereafter V04). The original dispersion of these spectrum templates is
0.4\AA, and a resolution of 1\AA\ full-width-half-maximum (FWHM), which are suitable for applying
to SDSS data (i.e. $\delta z/z=\delta V_{\rm SDSS}/c\sim 2\times 10^{-4}$, where $\delta z$ is the
accuracy of the measured $z$ and $\delta V_{\rm SDSS}\sim 70$\kms\ is the spectral resolution of SDSS).

We simultaneously obtain
velocity dispersion ($\sigma_*$) of galaxies and redshift correction ($\Delta z$)
from fitting the absorption line systems of the host galaxies by direct-pixel fitting method.
Bernardi et al. (2003) show in detail that the 4000-7000\AA\ is the best window for $\sigma_*$
in their SDSS sample of elliptical galaxies, however, this window could be not the best one for
$\Delta z$. We search reasonable windows for both $\sigma_*$ and $\Delta z$, and find the 3800-7000\AA\
is the best window since the mean values of $\chi^2$ are the minimum in our present sample. We find
$\langle\chi^2\rangle=1.2$ for 4000-7000\AA\ whereas $\langle\chi^2\rangle=1.1$ for 3800-7000\AA.
In the meanwhile, $\sigma_*$ values are consistent with each other for the 3800-7000\AA\ and
4000-7000\AA\ windows (we find that the difference of dispersion velocity,
$\langle\delta \sigma_*\rangle\approx 9\pm19{\rm km~s^{-1}}$, and
$\langle\delta V\rangle\approx 8\pm 11{\rm km~s^{-1}}$, both difference could be neglected, where
$\delta V=\delta(\Delta z)c$). Therefore, we select the 3800-7000\AA\ window mainly by justifying
$\langle\chi^2\rangle$ for the sample in this paper.

An example of the galactic spectra fitted with stellar templates is shown in Figure 2
{\it top-left} panel, and the $\chi^2$-distributions of the fittings of the sample in Figure 2
{\it top-right}. We find that the $\chi^2$ distribution of the fits in the restricted ranges
is better than the ones for the fits in the full spectral band. This means that the direct
pixel-fitting method only applies to determine the redshifts of galaxies through absorption lines,
rather than estimating the underlying continuum. Stellar velocity dispersions are measured mainly
through Ca {\sc ii} H+K and Mg {\sc i}b absorption lines. We have examined if there are some
sources with double systems of absorption lines, but no sources are found in the
sample\footnote{We also checked the 54 dual-cored galaxies given in Section 3, but they do not
show double systems of absorption lines either. The reasons for this phenomenon remain open.
It may result from the complicated situation of merging galaxies, in which the NLR dynamics
does not follow the merging galaxies. Clearly, it is worth of observing the 54 dual-cored
\ppgalaxies\ from the present sample through long-slit spectroscopy.}. Cosmological
redshifts of {\it all} emission-line galaxies are determined by this method in the present paper.

Using STARLIGHT (Cid Fernandes et al. 2005), we estimate the underlying continuum. This code
employs 45 stellar population templates (i.e. Base.BC03.N) which are composed of 3 metallicities
and 15 ages (Bruzual \& Charlot 2003, hereafter BC03). We make correction for Galactic extinction
(Cardelli et al. 1989) and redshift. Since the BC03 template spectra have a homogeneous wavelength
interval of 1\AA, we have to interpolate SDSS spectra to the same interval with the
templates\footnote{The sampling SDSS spectra of $\Delta \log \lambda=10^{-4}$ gives
$\delta\lambda\approx 1.15$\AA\ at 5007\AA, implying that the interpolated spectra are accurate
enough.}. STARLIGHT is used to fit the interpolated spectra and to get stellar absorption and
continuum spectra (SACS) of galaxies. Figure 2 {\it bottom-left} shows an example of the fit,
whereas the {\it right} panel shows the $\chi^2$-distribution of the fittings for the sample.
Comparing the $\chi^2$-distributions presented in Figure 2 {\it bottom-right}, we find that
applications of STARLIGHT to the whole spectral coverage are better than to the restricted
regions (see caption of Figure 2). Though STARLIGHT provides redshifts of
galaxies, its results seem to be less accurate than the ones obtained with the direct
pixel-fitting method (see Figure 2 {\it top-, bottom-right}). These comparisons show that
the STARLIGHT applies to estimate the underlying continuum whereas the direct pixel-fitting method to
determine the redshifts and stellar velocity dispersion.

We interpolate the SACS to the SDSS spectral points, and obtain pure emission-line spectra from SDSS
spectra by subtracting the interpolated SACS. Errors of the data points in the pure emission
line spectra are taken as the same as that in the SDSS original spectra. The pure emission-line spectra
are employed to determine the shifts and the fluxes of the double peaks.

Finally, we note that the subtraction of Balmer absorption lines could make some artificial broad
components though it is examined by $\chi^2$. In practice it turns out that the artificial components
are produced if the AGN components are relatively brighter than their hosts, especially in some type I
\ppagn\ with $\chi^2>1.5$. Fortunately, the fraction of these galaxies is less than $\sim 1\%$.

\subsection{Analysis of emission-line spectra}
The pure emission-line spectra may contain several kinetic components of emission lines as shown
in Figure 3: narrow lines (\nii, \oiii\, H$\alpha$, H$\beta$), \oiii\ wings (usually blueshifted, but
sometimes redshifted), broad Balmer emission lines. We fit the lines with several Gaussian components,
and classify them into different groups in light of kinetics. In each group, every Gaussian has the
same shift and width. The pure emission-line spectra are modeled by
(1) one or two groups for the NELs (narrow Gaussian group) corresponding to a single or double peaks;
(2) one group for \oiii$\lambda\lambda$ 4959, 5007 wings, respectively (\oiii\ wing group);
(3) one group for broad H$\alpha$ and H$\beta$ components (broad Balmer group). We stress
that the group components of each model are independent, in particular the widths,
the shifts and the fluxes of the \oiii\ wings are free in the fittings. Except for the
intensity ratios of \oiii $\lambda5007/\lambda4959=3$ and \nii\ $\lambda6584/\lambda6548=3$
(Osterbrock \& Ferland 2006), all line fluxes are set free.

There are eight models with different combinations of groups as shown in Table 2. We fit each
individual spectrum with all of the eight models and calculate $\chi^2$ in three wavelength bands:
$\chi^2_{\beta}$ for \oiii and H$\beta$ fitting in the rest-frame wavelength range 4770-5080\AA,
$\chi^2_{\alpha}$ for \nii and H$\alpha$ fitting in the range 6400-6700\AA,
and $\chi^2_t$ for fitting in the two ranges simultaneously. For galaxies with $z>0.39$, only
$\chi^2_{\beta}$ is available. After obtaining the $\chi^2$ of all the eight models, we
use {\it F}-test (Lupton 1993, Chap. 12.1)
to determine the best model for each sources through three quantities defined as
\begin{equation}
f_{\alpha}=\frac{\chi^2_{\alpha_1}/n_{\alpha_1}}{\chi^2_{\alpha_2}/n_{\alpha_2}},~~~
f_{\beta}=\frac{\chi^2_{\beta_1}/n_{\beta_1}}{\chi^2_{\beta_2}/n_{\beta_2}},~~~
f_t=\frac{\chi^2_{t_1}/n_{t_1}}{\chi^2_{t_2}/n_{t_2}}.
\end{equation}
where $f_{\alpha}$, $f_{\beta}$ and $f_t$ follow $F_{\alpha}$-, $F_{\beta}$- and $F_t$-distribution.
We take the minimum probability $P_{\rm min}=\min(P_{\alpha}, P_{\beta}, P_t)$,
where $P_{\alpha}, P_{\beta}, P_t$ are the probabilities defined as
\begin{equation}
P_{\alpha}=P(F>F_{\alpha}),~~~
P_{\beta}=P(F>F_{\beta}),~~~
P_t=P(F>F_t).
\end{equation}
These probabilities are used to check whether the model with more Gaussians can improve the fitting
in the corresponding wavelength range. If $1-P_{\rm min}>99.73\%$ ($3\sigma$), then the model with more
Gaussians is considered to be better. Such comparisons are done between the eight models, from the simplest
one to those with more and more Gaussians, until a model with more Gaussians can not
improve the fitting. We then reach the best model for all individual objects. After directly removing those
sources with a single Gaussian group from the best model, we are left with 118,871 sources with double narrow
Gaussian groups (model M2, M6, M7 or M8).

Since the reliability of $F$-test mainly depends on the S/N ratio, the intensity ratio of the narrow lines and
the confidence level, some sources with a single Gaussian group are incorrectly identified by the above $F$-test
as sources with red and blue components. We choose a low confidence level in order to avoid losing any candidates,
however, we proceed to remove those pseudo candidates according to the following empirical criterion.
We employ the conservative criteria as
\begin{equation}
0.03<F_{\rm r}/F_{\rm b}<30.0,
\end{equation}
\begin{equation}
\sigma_{\rm err, red}/\sigma_{\rm red}<0.2,~~~~\sigma_{\rm err, blue}/\sigma_{\rm blue}<0.2,
\end{equation}
where $F_{\rm r}$ and $F_{\rm b}$ are fluxes of red and blue components respectively,
$\sigma_{\rm red}$ and $\sigma_{\rm blue}$ are the corresponding velocity dispersions,
while $\sigma_{\rm err, red}$ and $\sigma_{\rm err, blue}$ are their errors. These empirical
criteria guarantee that the NELs do need two-Gaussians under the
signal-to-noise ratio of SDSS spectra. Otherwise, one of the two Gaussian components
will be overwhelmed with spectral noise (see Appendix A). With conditions (3) and (4), we
identify 107,745 sources which require
two Gaussians to fit their NELs. In the next subsection, we use the parameters
of the two Gaussians to obtain candidates with double-peaked and asym-NEL galaxies.

\subsection{Final sample of \ppgalaxies}
A profile composed of two narrow Gaussian components shows double peaks when the separation of the
two components is large enough. Using Monte-Carlo simulations (see Appendix A for details), we find
the following criteria useful for selecting candidates with double-peaked profiles:
\begin{equation}
\Delta V/{\rm FWHM_{min}}>0.8,~~{\rm or}~~\Delta V > 200 {\rm km~s^{-1}},
\end{equation}
where $\Delta V$ is the relative velocity shift between the two narrow Gaussian components,
$\rm FWHM_{\rm min}=\min(\rm FWHM_{\rm red}, FWHM_{\rm blue}$), $\rm FWHM_{\rm red}$ and
$\rm FWHM_{\rm blue}$ are FWHMs of two components respectively. After applying the criteria
presented in equation (5), we obtain 42,927 candidates of \ppgalaxies. As analyzed in Appendix A,
the above procedures efficiently exclude those single-peaked galaxies and make it possible to select
real \ppgalaxies\ through visual inspection of spectra, but they are mixed with asym-NEL galaxies.

We visually inspect the 42,927 galaxies, and find that there are 3,030 \ppgalaxies. As a by-product
of the present criteria, we find 12,582 asym-NEL galaxies. The \ppgalaxies\ are defined by the fact
that there is a clear ``trough" in the profiles of narrow emission lines. We select asym-NEL galaxies,
which do not show ``trough", and directly excluded those objects without strong asymmetric profiles.
As we show in the Appendix A, the double-peaked sources are complete by making use of the present criteria,
but asym-NEL sources are unavoidably selected. We point out that selection criteria of asym-NEL sources rule
out many potential sources (see Appendix A) so that the selected sample is not complete. However the
selected sources are still numerous up to 12,000, it is thus believed that statistical properties of
the sample are meaningful. As we show in subsequent sections, the \ppgalaxies\ and the asym-NEL galaxies
follow similar statistic properties. Table 3 and 4 list
all the \ppgalaxies\ and asym-NEL galaxies as well as the widths, the fluxes, the redshifts and
the blueshifts of the two peaks. The two tables are organized by object type (see Section 2.4).

\subsection{Classification of the sample}
Since the production of the red and blue peaks could be driven by different radiation mechanisms, we
diagnose the nature of each peak components through the BPT diagram
(Baldwin et al. 1981). Generally, the ratios of the narrow lines are probes of the nature of the
ionizing sources, which are either young stars in star-forming (SF) regions or active galactic nuclei.
If a galaxy has obvious broad components of Balmer lines (H$\alpha$ and H$\beta$), it should have at
least one type I AGN component. The criteria for the type I AGN follows Hao et al. (2005):
\begin{equation}
{\rm FWHM(H\alpha) > 1200 km~s^{-1}~~ and}~~ h_{\rm H\alpha, broad}/\bar{h}_{\rm H\alpha, narrow}>0.1,
\end{equation}
or
\begin{equation}
{\rm FWHM(H\alpha) > 2200 km~s^{-1}},
\end{equation}
where $\bar{h}_{\rm H\alpha, narrow}=(h_{\rm H\alpha, red}+h_{\rm H\alpha,blue})/2$ is the mean
height of specific flux of the blue ($h_{\rm H\alpha, blue}$) and red
($h_{\rm H\alpha, red}$) narrow component of H$\alpha$. FWHM(H$\alpha$) is the
full-width-half-maximum of the broad H$\alpha$ component.

Figure 4 shows the BPT diagram of the red and the blue components of the present sample.
The demarcation described in equation 1 of Kauffmann et al. (2003) is used to distinguish AGN
and SF components in the BPT diagram. Several combinations are displayed in Figure 4: two type
I AGN\footnote{In practice, we directly pick up those objects with broad emission lines as type
I AGN as given by conditions (6) or (7).}; one type I and one
type II AGN; one type I AGN and one SF components (we call all the three kinds as `type I');
two type II AGN (`2-type II'); two SF components (`2-SF'); one type II AGN and one SF components
(`type II + SF'). In practice, the quality of the current SDSS spectra does not allow us to distinguish
the two cases: 2 type I AGN and type I+type II AGN. We generally classify them as type I \ppagn.
Galaxies with weak \oiii and H$\beta$ (i.e. the amplitudes of both are lower
than 3 times the standard deviation of the local continuum), and those with $z>0.39$ (H$\alpha$
and \nii do not lie within the SDSS spectral coverage) can not be classified by the BPT diagram,
are noted as ``unknown-type" (hereafter ``unknown").
We list numbers of each type of objects found in the present paper in Figure 1. Figures 5
and 6 show examples of the five types with images and
spectra (\ppgalaxies\ and asym-NEL galaxies).

Finally, we separate sources with strong \oiii\ wings given in Table 5. They show interesting properties,
which can be used to test outflow models, however, detailed discussions on the \oiii-winged \ppgalaxies\
are beyond the scope of this paper.

\subsection{Comparison with previous samples}
The present paper provides much more objects than previous samples (the total is $\sim 190$
\ppagn\ from W09 and L10) and covers {\it all} the known \ppagn\ from SDSS galaxy sample (DR7). The
main reason for the dramatic increase of the number of objects is the selection criterion for AGN. The
previous studies use the Kewley et al. (2001) criterion to select AGN in a more strict way whereas
this work makes use of the empirical criterion by Kauffmann et al. (2003).
We prefer this empirical criterion to
classify the AGN and SF galaxies since sources located in the region between the Kauffmann
and the Kewley et al. criteria are generally considered as AGN.
In order to show the influence of the different criteria on the sizes of the resultant samples,
we start from the $3,030$ \ppgalaxies\ selected by the present paper.
Using the MPA measurements of the \ppgalaxies, 222 \ppagn\ are left by applying the Kewley
et al. criterion. Therefore, only 32 objects do not appear in the previous studies. These 32 objects
are due to the EW(\oiii) selection criterion: EW(\oiii) is larger than $\rm 5\AA$ in W09,
larger than $\rm 4\AA$ in L10, but larger than $\rm 3\AA$ in our work.

\section{\ppgalaxies: SDSS images}
After selecting the two samples of \ppgalaxies\ and asym-NEL galaxies, we
visually inspect the SDSS images of the present samples (15,612 galaxies) to find dual-cored
galaxies. Sub-structures of spiral
galaxies could be regarded as dual-cored galaxies, but in practice it turns out that there
are only a few of such cases in the present sample. The proper identification of substructures
will require observing the spectrum of the second core in the future.
If the dual-cored galaxies are due to projection effects (i.e. background or foreground galaxies),
these pseudo dual-cored galaxies can be excluded if the redshift differences of the absorption lines
is larger than 0.0017 (e.g. Patton et al. 2000). Visual inspection of potential dual-cores
is limited by the SDSS image resolution ($\sim 1.4^{\prime\prime}$). We find 54 dual-cored
$pp$-galaxies and asym-NEL galaxies from SDSS images, which are listed in Table 6 and Figure 11 of
Appendix B. Among these galaxies, there are 14 2-type II AGN, 8 2-SF galaxies, 12 type
II+SF galaxies, and 20 unknown-type galaxies (or noted by ``unknown"). The 54 galaxies are the
largest sample of dual-cored galaxies with double-peaked profiles from SDSS as far as we know.

We stress here that the 54 dual-cored \ppgalaxies\ and asym-NEL galaxies do not fully
rule out the potential roles of the co-existing outflows. The observed double-peaked components
could be contributed by the outflows such as Mrk 78 (Fisher et al. 2011). However, the current data
does not allow us to separate the potential contributions of outflows. High quality observations
of each cores should be made for more details of the dual-cored \ppgalaxies. On the other
hand, IFU observations are necessary to study the details of dynamics of merging galaxies.

Besides the 54 dual-cored \ppgalaxies, visual inspection yields a by-product sample of 261
dual-cored \ppgalaxies\ and asym-NEL galaxies, for which the second cores are beyond the
SDSS $3^{\prime\prime}$ fibers. We also check if the second cores have been observed by SDSS
spectroscopic observation, and find 45 second cores with SDSS spectra. If the redshift difference
between the primary and second cores is $\Delta z>0.0017$, they could be
irrelevant (Patton et al. 2000; Rogers et al. 2009; Lin et al. 2010). We exclude 6 of this kind
of sources as foreground or background galaxies, or stars. The left 255 dual-cored galaxies in
our samples are listed in Table 7 and Figure 12 of Appendix C. We find that there are two 
interesting \ppgalaxies\ of which
the second cores have asymmetric profiles. They are SDSS J001139.74-002827.8 accompanied by
J001139.98-002826.1 (asym-NEL) and SDSS J115610.40+045935.0 by J115610.77+045943.4 (asym-NEL).
Further observations (e.g. IFU) are needed to identify the origin of
the emission-line profiles and the dynamics of the merging system.

We have checked FIRST radio images of the present \ppgalaxies\ sample. It turns out that they show very
interesting morphologies with different sizes. The results will be carried out in the second paper of this
series. Though VLBA images of 87 \ppagn\ do not show evidence for dual cores (Tingay \& Wayth 2011), a
systematical VLBA survey is necessary among the present sample of the \ppgalaxies.

\section{Properties of \ppgalaxies\ and asym-NEL galaxies}
With the measurements of profile parameters, we can investigate correlations among them to explore the
nature of the double peaks. However, we focus on the basic properties of the two samples of double-peaked
narrow lines in this paper.

\subsection{Redshift distributions}
Figure 7{\it a,b} show the redshift distributions of different kinds of \ppgalaxies\ and asym-NEL
galaxies. The samples have a redshift range from 0 to 0.6, but $99\%$ of the sources have $z<0.3$. No
significant differences among sub-samples have been found, implying the sub-samples are
homogeneous in redshifts. There are no significant difference between $pp$-galaxies and asym-NEL
galaxies. Figure 7{\it c} shows the redshift distribution of the 54 dual-cored \ppgalaxies.
They are homogeneously distributed between $0.04\le z\le 0.2$ with a mean value of $\langle z\rangle=0.12$.

It should be mentioned that dual-cored galaxies ($\sim 310$) found by Galaxy Zoo project (Darg et al.
2010, however they have not published their catalog so far.) are selected from $0.005\le z\le 0.1$
galaxy sample of SDSS DR6. This image-selected sample
of dual-cored galaxies is seriously suffering from the SDSS spatial resolution
($\Delta \theta\sim 1.4^{\prime\prime}$). The present sample is based on spectral profiles,
and may cover some dual-cored candidates with more closer distance than Darg et al. sample, but
should be identified through images with higher resolution. For example,
NIR imaging observations with higher angular resolution are useful for identifying those galaxies
with closer two cores (e.g. Fu et al. 2011, Shen et al. 2011, and Rosario et al. 2011).
From Figure 7{\it c}, it is found that the dual-cored galaxies are not special populations in
$z-$distributions.

\subsection{$\Delta V$ distributions}
After the host galaxy redshift is determined, the relative shifts of the blue and red peaks
($\Delta \lambda_{\rm b}$ and $\Delta \lambda_{\rm r}$) of {\it pp}-galaxies at \oiii wavelength
can be obtained. Here $\Delta \lambda_{\rm b}=\lambda_{\rm b}-\lambda_0$,
$\Delta \lambda_{\rm r}=\lambda_{\rm r}-\lambda_0$, $\lambda_{\rm b}$ and $\lambda_{\rm r}$ are
the wavelengths of blue and red components respectively, and $\lambda_0$ is the wavelength of \oiii
$\lambda 5007$ in the rest frame. The distributions of blueshift and redshift velocities are shown
in Figure 8 {\it top} panels ($a_1-e_1$), where the averaged velocities are indicated. We find the
following properties from the distributions: 1) $\langle\dvr\rangle$ is slightly larger than
$\langle\dvb\rangle$ by $\sim 60$\kms\ for type I AGN whereas $\langle\dvr\rangle$ and
$\langle\dvb\rangle$ have similar distributions in the other four types of {\it pp}-galaxies;
2)the shift ratio $\langle {\rm log} ~v \rangle = \langle {\rm log} ~|\dvr/\dvb| \rangle \simeq 0$
except for type I \ppagn; 3) 2-SF \ppgalaxies\ generally have quite narrow distributions of
redshifts and blueshifts whereas type I $pp$-AGN have relatively broad distributions.

$\Delta V_{\rm r,b}$ distributions of the asym-NEL galaxies are shown in Figure 8 {\it bottom}
panels ($a_2-e_2$). From the five panels, we find: 1) $\langle\dvr\rangle \sim \langle\dvb\rangle$;
2) the shift ratio $\langle {\rm log} ~v \rangle \simeq 0$ for all the five kinds of galaxies;
3) the distribution properties are similar to $pp$-galaxies except for smaller velocity separations.

\subsection{Distribution of flux ratios}
Distributions of two integrated flux ratios of H$\alpha$ are shown in Figure 9{\it a}, and that of \oiii\ in
Figure 9{\it b}. We find that the sample mainly distributes within $[0.1,10]$, which is a selecting
bias. If flux ratios are too small or too large, the component with larger flux is dominating,
and another component is so weak that it requires to be identified by higher S/N observations.

From Figure 9{\it a}, $F_{\rm H\alpha}^{\rm r}/F_{\rm H\alpha}^{\rm b}$ generally has narrow
log-normal distributions for both \ppgalaxies\ and asym-NEL galaxies. Except for type 1 \ppagn\
with broader
$F_{\rm H\alpha}^{\rm r}/F_{\rm H\alpha}^{\rm b}$ distribution, the other types are very similar.
Figure 9{\it b} shows $F_{\rm [O~III]}^{\rm r}/F_{\rm [O~III]}^{\rm b}$ distributions, similar to
that of $F_{\rm H\alpha}^{\rm r}/F_{\rm H\alpha}^{\rm b}$, but the former is relatively broader than
the later. It is interesting to note that both $F_{\rm H\alpha}^{\rm r}/F_{\rm H\alpha}^{\rm b}$ and
$F_{\rm [O~III]}^{\rm r}/F_{\rm [O~III]}^{\rm b}$ distributions in 2-SF are the narrowest among the five
kinds of \ppgalaxies\ and asym-NEL galaxies. This implies that there are undergoing processes with symmetric
emission in 2-SF galaxies, that is to say, they could have either symmetric bi-polar outflows or circular
rotation-supported NLR, or equal mass dual-cores.
Finally, we find that the distributions of the flux ratios in \ppgalaxies\ and asym-NEL galaxies are similar.

\section{Conclusions}
We have developed a pipeline to automatically select candidates of double-peaked galaxies of spectra.
Combining with visual inspection of spectra, we set up the largest sample of double-peaked galaxies to
date, which is composed of double-peaked narrow emission-line galaxies (3,030). We also build a sample
of galaxies (12,582) with asymmetric and top-flat profiles from the SDSS DR7 galaxy sample. The \ppgalaxies\
and the asym-NEL galaxies show very similar statistical properties, but velocity separations are significantly
different. We find that double-peaked components are divided into 5 kinds, namely, type I \ppagn, 2-type II
\ppagn, (AGN+SF) \ppgalaxies, 2-SF and unknown-type
objects depending on the Balmer line width and the Baldwin-Phillips-Terlevich diagram. The \ppgalaxies\
sample covers: 81 type I \ppagn, 837 2-type II \ppagn, 400 type II+SF \ppgalaxies, 708 2-SF and 1,004 unknown-type
objects. We visually inspect the SDSS images of the sample and find 54 galaxies with dual-cores, of which
14 2-type II \ppagn, 8 2-SF, 12 type II+SF \ppgalaxies\, and 20 unknown-type objects.

We would draw the following conclusions in light of the present sample:
\begin{itemize}
\item The fractions of \ppgalaxies\ and asym-NEL galaxies to emission-line galaxies are about $1.0\%$ and
      $3.6\%$, respectively, and \ppagn\ (including type I, 2-type II, type II +SF \ppgalaxies) and
      asym-NEL AGN are about $0.4\%$ and $1.5\%$, respectively.

\item From the two samples, we find 54 dual cored galaxies with
    projected separation smaller than $3^{\pp}$ and 255 with that larger than $3^{\pp}$.
    Observations with higher spatial resolution or in X-ray or radio bands are needed to identify
    dual-cores from images of rest objects of the present samples, in particular, the sub-samples of type
    I and AGN+SF \ppgalaxies.
\end{itemize}

Future observations with for example the {\it LAMOST/Guoshoujing} (a 4-m telescope with
4,000 fibers) through images and spectra of the two
components are needed to understand the nature of the $pp$-galaxies. The large sample can be used to
study the dynamical processes of merging galaxies, the triggering mechanisms of AGN activity, outflows
and rotation-supported disk of narrow line regions.

\acknowledgements{We are very grateful to the anonymous referee for a large number of helpful comments
and suggestions in careful reports, which improved the manuscript. JMW thanks David Valls-Gaubald
for carefully reading the manuscript and suggestions. We appreciate the stimulating
discussions among the members of IHEP AGN group, especially, Long Di, Han-Qin Gao and Zhen Zhang, who
visually re-checked the images of 15,000 double-peaked galaxies, and Yan-Rong Li for useful discussions
on star formation driven by tidal interaction. The research is supported by NSFC-10733010,
10821061, 10903008 and 11133006, and 973 project (2009CB824800). SZ thanks support of NSFC-11073021
and 11133002. Funding for the SDSS and SDSS-II has been provided by the Alfred P. Sloan Foundation,
the Participating Institutions, the National Science Foundation, the U.S. Department of Energy, the
National Aeronautics and Space Administration, the Japanese Monbukagakusho, the Max Planck Society,
and the Higher Education
Funding Council for England. The SDSS Web Site is http://www.sdss.org/. The SDSS is managed by the
Astrophysical Research Consortium for the Participating Institutions. The Participating Institutions
are the American Museum of Natural History, Astrophysical Institute Potsdam, University of Basel,
University of Cambridge, Case Western Reserve University, University of Chicago, Drexel University,
Fermilab, the Institute for Advanced Study, the Japan Participation Group, Johns Hopkins University,
the Joint Institute for Nuclear Astrophysics, the Kavli Institute for Particle Astrophysics and Cosmology,
the Korean Scientist Group, the Chinese Academy of Sciences (LAMOST), Los Alamos National Laboratory, the
Max-Planck-Institute for Astronomy (MPIA), the Max-Planck-Institute for Astrophysics (MPA), New Mexico
State University, Ohio State University, University of Pittsburgh, University of Portsmouth, Princeton
University, the United States Naval Observatory, and the University of Washington.}


\clearpage

\begin{deluxetable}{llll}
\tablecolumns{4}
\tablewidth{0pc}
\tablecaption{Summary of SDSS samples of double-peaked AGN}
\tabletypesize{\footnotesize}
\tablehead{
\colhead{Sample} & \colhead{Selection methods} & \colhead{Object number}
& \colhead{Notes}
}
\startdata
W09     & $r<17.7$, $z\le 0.15$, EW(\oiii)$>5$\AA, $0.1\le F_{\rm red}/F_{\rm blue}\le 10$ &87& only type II AGN \\
L10     & S/N$>5$, \oiii$>5\sigma$, EW(\oiii)$\ge 4$\AA, \oiii/H$\beta>3$ &146+21 &only type II AGN+QSO \\
S10 & $0.1\le z\le 0.7$ QSOs & 148 & type I, II \\
Others  & individual objects                         & 8     &\ppagn         \\
This work & see Figure 1                             &3,030+12,582 &covers all previous galaxies\\ \hline
\multicolumn{4}{l}{\ppagn\ observed with near infrared (NIR) higher spatial resolution, long slit
spectroscopy and Integral Field Unit (IFU)}\\ \hline
Fu11   & from W09+L10+Smith10; NIR image     & 17+33  &mergers: 8/17+8/33\\
Shen11  & from L10; NIR image+long slit spectroscopy & 31     &mergers: 5        \\
R11     & from Smith10; NIR image                    & 12     &mergers: 6        \\
Fu12   & from W09+L10+Smith10; IFU     & 42     &dual AGN: 2       \\
C11     & from W09+L10; long slit spectroscopy     & 81     &dual AGN candidates: 17
\enddata
\tablecomments{References: C11: Comerford et al. (2011); Fu11: Fu et al. (2011); Fu12: Fu et al. (2012);
L10: Liu et al. (2010a); R11: Rosario et al. (2011); S10: Smith et al. (2010); Shen11: Shen et al. (2011);
W09: Wang et al. (2009);\\
--- Others: SDSS J09527.62+255257.2 (McGurk et al. 2011), COSMOS J100043.15+020637.2 (Comerford et al.
2009b), SDSS J104807.74+005543.5 (Zhou et al. 2004), SDSS J131642.90+175332.5 (Xu \& Komossa 2009),
EGSD2 J141550.8+520929 (Comerford et al. 2009a), EGSD2 J142033.66+525917.5 (Gerke et al. 2007),
SDSS J142507.32+323137.4 (Peng et al. 2011), CXOXBJ142607.6+353351 (Barrows et al. 2012).\\
--- In Fu11, 17 type I AGN and 33 type II AGN are imaged by Keck II LGSAO in the NIR, in which 8 type I
AGN and 8 type II AGN are found to be mergers.\\
--- 3,030 \ppgalaxies\ and 12,582 asym-NEL galaxies are found in this work.\\
--- We would like to stress the different criterion to select \ppagn\ in W09 and L10 from the present
paper. The Kauffmann et al. (2003) criterion (empirical one) is used in the present paper whereas the Kewley
et al. (2001) criterion was used in W09 and L10.
}
\end{deluxetable}

\begin{deluxetable}{ccccc}
\tablecolumns{5}
\tablewidth{0pc}
\tablecaption{Eight fitting models of different groups of emission lines}
\tabletypesize{\footnotesize}
\tablehead{
\colhead{model$\backslash$group} & \colhead{narrow Gaussian}   & \colhead{narrow Gaussian}    & \colhead{\oiii wing}
 & \colhead{broad Balmer}
}
\startdata
M1 & $\surd$ &          &          &             \\
M2 & $\surd$ & $\surd$ &          &             \\
M3 & $\surd$ &          & $\surd$ &             \\
M4 & $\surd$ &          &          & $\surd$      \\
M5 & $\surd$ &          & $\surd$ & $\surd$     \\
M6 & $\surd$ & $\surd$ &          & $\surd$       \\
M7 & $\surd$ & $\surd$ & $\surd$ &             \\
M8 & $\surd$ & $\surd$ &$\surd$  & $\surd$       \\
\enddata
\tablecomments{
M1-M8 are the eight fitting models which are different combinations of four emission-line groups: two
narrow Gaussian groups, one \oiii wing group, and one broad Balmer group.
}
\end{deluxetable}

\begin{deluxetable}{lccccccccccccccc}
\rotate

\tabletypesize{\tiny}
\tablecolumns{16}
\tablewidth{0pc}
\tablecaption{Emission-line galaxies with double-peaked NELs}
\tablehead{
\colhead{SDSS name} & \colhead{mjd}   & \colhead{plate}    & \colhead{fiber}   & \colhead{$z$} & \colhead{$r$} & \colhead{$\sigma_{\rm *}$}  &
\colhead{$\Delta V_{\rm b}$} & \colhead{$\Delta V_{\rm r}$} & \colhead{$\sigma_{\rm b}$}    & \colhead{$\sigma_{\rm r}$} &
\colhead{$F_{\rm [OIII]}^{\rm b}$}    & \colhead{$F_{\rm [OIII]}^{\rm r}$}   & \colhead{$F_{\rm H\alpha}^{\rm b}$}    &
\colhead{$F_{\rm H\alpha}^{\rm r}$}  & \colhead{Type}\\
\colhead{(1)} & \colhead{(2)}   & \colhead{(3)}    & \colhead{(4)}   & \colhead{(5)}   & \colhead{(6)} &
\colhead{(7)} & \colhead{(8)}    & \colhead{(9)}   & \colhead{(10)}    &
\colhead{(11)} & \colhead{(12)}   & \colhead{(13)} & \colhead{(14)}  & \colhead{(15)} & \colhead{(16)}\\
}
\startdata
\multicolumn{16}{c}{type I} \\ \hline
J011935.64-102613.2  & 52163  &  661  & 285  &   0.12479$\pm$  0.00004 & 17.32 &    138$\pm$  10 &    -76$\pm$   8 &    235$\pm$   4 &    161$\pm$   5 &    125$\pm$   2 &    764$\pm$  38 &    831$\pm$  37 &    343$\pm$   9 &    250$\pm$   9 &   1 \\
J040001.59-065254.1  & 51908  &  464  & 104  &   0.17003$\pm$  0.00003 & 16.87 &    227$\pm$   7 &   -144$\pm$   6 &    229$\pm$   3 &    183$\pm$   5 &    142$\pm$   2 &    734$\pm$  30 &    813$\pm$  29 &    243$\pm$   7 &    288$\pm$   8 &   1 \\
J074156.55+294135.6  & 52663  &  889  & 271  &   0.12115$\pm$  0.00003 & 16.77 &    119$\pm$   8 &   -110$\pm$   5 &    103$\pm$   5 &     92$\pm$   4 &     77$\pm$   3 &     28$\pm$   2 &     18$\pm$   2 &    178$\pm$  12 &    135$\pm$  12 &   1 \\
J075017.49+270304.1  & 52316  &  858  & 437  &   0.14104$\pm$  0.00003 & 16.84 &    150$\pm$   9 &    -90$\pm$   9 &    133$\pm$   8 &    111$\pm$   5 &     94$\pm$   4 &     52$\pm$   4 &     21$\pm$   3 &    307$\pm$  16 &    234$\pm$  21 &   1 \\
\cutinhead{2-type II}
J000249.07+004504.8  & 51793  &  388  & 345  &   0.08650$\pm$  0.00004 & 16.08 &    221$\pm$  12 &   -248$\pm$   3 &    238$\pm$   2 &    162$\pm$   3 &    158$\pm$   2 &    444$\pm$  12 &    639$\pm$  21 &    446$\pm$  15 &    551$\pm$  15 &   2 \\
J000656.85+154847.9  & 52251  &  751  & 504  &   0.12498$\pm$  0.00002 & 16.92 &    152$\pm$   6 &   -193$\pm$   7 &    174$\pm$   6 &    156$\pm$   4 &    151$\pm$   4 &    493$\pm$  22 &    502$\pm$  22 &    188$\pm$   6 &    225$\pm$   6 &   2 \\
J001630.41-003801.6  & 51795  &  389  & 163  &   0.06346$\pm$  0.00003 & 16.20 &    169$\pm$   8 &   -135$\pm$   4 &    176$\pm$   4 &    125$\pm$   3 &    119$\pm$   3 &     59$\pm$   3 &     38$\pm$   2 &    516$\pm$  21 &    524$\pm$  26 &   2 \\
J002202.11-100744.1  & 52145  &  653  & 227  &   0.24761$\pm$  0.00006 & 17.70 &    254$\pm$  15 &   -283$\pm$  19 &    131$\pm$  22 &    136$\pm$  15 &    205$\pm$  20 &     27$\pm$   4 &     41$\pm$   4 &      9$\pm$   4 &     48$\pm$   5 &   2 \\
\cutinhead{type II+SF}
J000712.22+155207.9  & 52251  &  751  & 578  &   0.16131$\pm$  0.00005 & 17.52 &    230$\pm$  15 &   -154$\pm$  12 &    186$\pm$  14 &    113$\pm$  11 &    135$\pm$  12 &      3$\pm$   1 &      4$\pm$   1 &     80$\pm$  28 &     94$\pm$  27 &   3 \\
J001050.52-103246.8  & 52141  &  651  & 108  &   0.15711$\pm$  0.00021 & 17.41 &    396$\pm$  56 &   -316$\pm$   2 &    -44$\pm$   8 &     77$\pm$   4 &    204$\pm$   6 &     33$\pm$   3 &     87$\pm$   3 &    128$\pm$  11 &    423$\pm$  19 &   3 \\
J002418.69-003426.1  & 53734  & 1542  &  63  &   0.34060$\pm$  0.00016 & 19.00 &    190$\pm$  43 &    -40$\pm$   7 &    218$\pm$   2 &    151$\pm$   5 &     87$\pm$   2 &     57$\pm$   2 &     33$\pm$   2 &    210$\pm$   6 &    156$\pm$   5 &   3 \\
J005104.37+152101.2  & 51871  &  420  & 436  &   0.15528$\pm$  0.00005 & 17.82 &    180$\pm$  15 &   -125$\pm$   4 &    171$\pm$   8 &     99$\pm$   3 &    114$\pm$   6 &     10$\pm$   0 &     13$\pm$   1 &    165$\pm$  12 &    116$\pm$  15 &   3 \\
\cutinhead{2-SF}
J000511.71+153846.2  & 52251  &  751  & 427  &   0.05356$\pm$  0.00003 & 15.16 &    182$\pm$   7 &   -110$\pm$   5 &    183$\pm$   6 &    127$\pm$   3 &    108$\pm$   4 &     30$\pm$   1 &     21$\pm$   2 &    713$\pm$  36 &    450$\pm$  35 &   4 \\
J000728.10-003617.1  & 52559  &  669  &  93  &   0.32712$\pm$  0.00008 & 19.07 &    135$\pm$  23 &   -123$\pm$  13 &    127$\pm$  12 &     95$\pm$  12 &     79$\pm$  10 &      0$\pm$   0 &      0$\pm$   0 &     30$\pm$  15 &     24$\pm$  11 &   4 \\
J001139.74-002827.8  & 52913  & 1089  & 205  &   0.05699$\pm$  0.00003 &  0.00 &     93$\pm$   8 &      9$\pm$   1 &    429$\pm$   3 &     91$\pm$   1 &     93$\pm$   3 &     18$\pm$   1 &     33$\pm$   2 &    403$\pm$  24 &     80$\pm$   6 &   4 \\
J002010.68-004852.8  & 51900  &  390  & 226  &   0.14129$\pm$  0.00002 & 16.07 &    181$\pm$   6 &   -118$\pm$   7 &    127$\pm$   7 &     94$\pm$   6 &     91$\pm$   5 &      3$\pm$   0 &      4$\pm$   0 &     73$\pm$  14 &     69$\pm$  11 &   4 \\
\cutinhead{unknown-type}
J000218.20-110139.5  & 52143  &  650  &  81  &   0.11959$\pm$  0.00002 & 16.76 &    172$\pm$   6 &   -118$\pm$   9 &    125$\pm$  14 &     79$\pm$   8 &    113$\pm$  12 &     12$\pm$   3 &     16$\pm$   3 &     72$\pm$  14 &     86$\pm$  13 &   5 \\
J000503.84+010629.6  & 52559  &  669  & 364  &   0.23857$\pm$  0.00007 & 18.79 &    151$\pm$  20 &    -86$\pm$  19 &    199$\pm$   9 &    122$\pm$  15 &    103$\pm$   7 &      2$\pm$   0 &      2$\pm$   0 &     43$\pm$   8 &     64$\pm$   9 &   5 \\
J001552.74-085430.9  & 52138  &  652  & 409  &   0.13692$\pm$  0.00008 & 17.39 &    281$\pm$  22 &   -181$\pm$  17 &    239$\pm$   9 &    216$\pm$  16 &    116$\pm$   8 &      5$\pm$   0 &     12$\pm$   2 &    132$\pm$  25 &     75$\pm$  16 &   5 \\
J001746.35-005052.7  & 52518  &  687  &  45  &   0.19020$\pm$  0.00004 & 17.82 &    170$\pm$  11 &   -123$\pm$  15 &    150$\pm$  16 &     88$\pm$  12 &    103$\pm$  17 &      1$\pm$   0 &      3$\pm$   1 &     28$\pm$  10 &     28$\pm$   9 &   5 \\
\enddata
\tablecomments{Table 3 is available in its entirety via the link to the machine-readable version above.\\
Column 1: SDSS DR7 designation hhmmss.ss+ddmmss.s (J2000.0). Column 2: MJD of spectroscopic observation.
Column 3: Plate of spectroscopic observation. Column 4: Fiber of spectroscopic observation.
Column 5: Redshift of spectroscopic observation.
Column 6: Magnitude in $r$-band.
Column 7: Velocity dispersion of galaxy absorption lines in units of km/s.
Column 8 and 9: the Doppler blueshifts and redshifts of the blue and red components of NELs in units of km/s, respectively.
Column 10 and 11 are the corresponding velocity dispersion in units of km/s.
Column 12 and 13 are the corresponding \oiii flux in units of $\rm 10^{-17} erg ~ s^{-1} ~ cm^{-2}$.
Column 14 and 15 are the corresponding H$\alpha$ flux in units of $\rm 10^{-17} erg ~ s^{-1} ~ cm^{-2}$.
Column 16: The type of the $pp$-galaxy: 1 standards for type I, 2 for 2-type II, 3 for type II + SF, 4 for 2-SF, and 5
for 'unknown-type'.\\
$-$ The machine-readable version of this table gives more effective numbers from Column (7) to (15).
}
\end{deluxetable}

\newsavebox{\tablebox}

\begin{deluxetable}{lccccccccccccccc}
\rotate
\tabletypesize{\tiny}
\tablecolumns{16}
\tablewidth{0pc}
\tablecaption{Emission-line galaxies with asymmetric and top-flat NELs}
\tablehead{
\colhead{SDSS name} & \colhead{mjd}   & \colhead{plate}    & \colhead{fiber}   & \colhead{$z$} & \colhead{$r$} & \colhead{$\sigma_{\rm *}$}  &
\colhead{$\Delta V_{\rm b}$} & \colhead{$\Delta V_{\rm r}$} & \colhead{$\sigma_{\rm b}$}    & \colhead{$\sigma_{\rm r}$} &
\colhead{$F_{\rm [OIII]}^{\rm b}$}    & \colhead{$F_{\rm [OIII]}^{\rm r}$}   & \colhead{$F_{\rm H\alpha}^{\rm b}$}    &
\colhead{$F_{\rm H\alpha}^{\rm r}$}  & \colhead{Type}\\
\colhead{(1)} & \colhead{(2)}   & \colhead{(3)}    & \colhead{(4)}   & \colhead{(5)}   & \colhead{(6)} &
\colhead{(7)} & \colhead{(8)}    & \colhead{(9)}   & \colhead{(10)}    &
\colhead{(11)} & \colhead{(12)}   & \colhead{(13)} & \colhead{(14)}  & \colhead{(15)} & \colhead{(16)}\\
}
\startdata
\multicolumn{16}{c}{type I} \\ \hline
J011935.64-102613.2  & 52163  &  661  & 285  &   0.12479$\pm$  0.00004 & 17.32 &    139$\pm$  11 &    -77$\pm$   8 &    235$\pm$   5 &    161$\pm$   5 &    125$\pm$   3 &    765$\pm$  39 &    832$\pm$  37 &    344$\pm$   9 &    250$\pm$   9 &   1 \\
J040001.59-065254.1  & 51908  &  464  & 104  &   0.17003$\pm$  0.00003 & 16.87 &    227$\pm$   8 &   -144$\pm$   7 &    229$\pm$   4 &    183$\pm$   5 &    143$\pm$   3 &    735$\pm$  30 &    814$\pm$  30 &    243$\pm$   8 &    288$\pm$   9 &   1 \\
J074156.55+294135.6  & 52663  &  889  & 271  &   0.12115$\pm$  0.00003 & 16.77 &    120$\pm$   8 &   -111$\pm$   6 &    103$\pm$   6 &     92$\pm$   4 &     78$\pm$   4 &     29$\pm$   2 &     19$\pm$   2 &    178$\pm$  12 &    136$\pm$  12 &   1 \\
J075017.49+270304.1  & 52316  &  858  & 437  &   0.14104$\pm$  0.00003 & 16.84 &    151$\pm$  10 &    -90$\pm$   9 &    133$\pm$   9 &    112$\pm$   5 &     95$\pm$   5 &     52$\pm$   4 &     22$\pm$   3 &    307$\pm$  17 &    235$\pm$  22 &   1 \\
\cutinhead{2-type II}
J000249.07+004504.8  & 51793  &  388  & 345  &   0.08650$\pm$  0.00004 & 16.08 &    222$\pm$  12 &   -248$\pm$   3 &    239$\pm$   3 &    163$\pm$   3 &    159$\pm$   3 &    444$\pm$  13 &    639$\pm$  22 &    447$\pm$  16 &    551$\pm$  16 &   2 \\
J000656.85+154847.9  & 52251  &  751  & 504  &   0.12498$\pm$  0.00002 & 16.92 &    152$\pm$   6 &   -194$\pm$   8 &    174$\pm$   7 &    157$\pm$   5 &    152$\pm$   4 &    493$\pm$  22 &    502$\pm$  22 &    189$\pm$   6 &    226$\pm$   6 &   2 \\
J001630.41-003801.6  & 51795  &  389  & 163  &   0.06346$\pm$  0.00003 & 16.20 &    170$\pm$   8 &   -135$\pm$   5 &    177$\pm$   5 &    126$\pm$   4 &    119$\pm$   3 &     59$\pm$   3 &     38$\pm$   3 &    517$\pm$  21 &    524$\pm$  26 &   2 \\
J002202.11-100744.1  & 52145  &  653  & 227  &   0.24761$\pm$  0.00006 & 17.70 &    254$\pm$  16 &   -283$\pm$  19 &    132$\pm$  23 &    136$\pm$  16 &    206$\pm$  20 &     28$\pm$   5 &     41$\pm$   5 &      9$\pm$   4 &     49$\pm$   5 &   2 \\
\cutinhead{type II+SF}
J000712.22+155207.9  & 52251  &  751  & 578  &   0.16131$\pm$  0.00005 & 17.52 &    230$\pm$  15 &   -155$\pm$  12 &    187$\pm$  15 &    114$\pm$  11 &    136$\pm$  13 &      4$\pm$   2 &      4$\pm$   1 &     81$\pm$  29 &     94$\pm$  27 &   3 \\
J001050.52-103246.8  & 52141  &  651  & 108  &   0.15711$\pm$  0.00021 & 17.41 &    396$\pm$  56 &   -316$\pm$   3 &    -45$\pm$   8 &     77$\pm$   4 &    205$\pm$   6 &     33$\pm$   4 &     87$\pm$   4 &    128$\pm$  11 &    423$\pm$  19 &   3 \\
J002418.69-003426.1  & 53734  & 1542  &  63  &   0.34060$\pm$  0.00016 & 19.00 &    190$\pm$  44 &    -41$\pm$   8 &    218$\pm$   3 &    152$\pm$   5 &     87$\pm$   3 &     58$\pm$   2 &     33$\pm$   2 &    211$\pm$   6 &    157$\pm$   6 &   3 \\
J005104.37+152101.2  & 51871  &  420  & 436  &   0.15528$\pm$  0.00005 & 17.82 &    180$\pm$  15 &   -125$\pm$   4 &    172$\pm$   8 &     99$\pm$   4 &    115$\pm$   7 &     11$\pm$   1 &     13$\pm$   2 &    165$\pm$  13 &    117$\pm$  16 &   3 \\
\cutinhead{2-SF}
J000511.71+153846.2  & 52251  &  751  & 427  &   0.05356$\pm$  0.00003 & 15.16 &    182$\pm$   7 &   -111$\pm$   5 &    183$\pm$   6 &    127$\pm$   4 &    109$\pm$   4 &     30$\pm$   2 &     21$\pm$   2 &    714$\pm$  36 &    450$\pm$  36 &   4 \\
J000728.10-003617.1  & 52559  &  669  &  93  &   0.32712$\pm$  0.00008 & 19.07 &    135$\pm$  24 &   -123$\pm$  13 &    128$\pm$  12 &     96$\pm$  13 &     79$\pm$  11 &      1$\pm$   0 &      1$\pm$   0 &     30$\pm$  15 &     24$\pm$  11 &   4 \\
J001139.74-002827.8  & 52913  & 1089  & 205  &   0.05699$\pm$  0.00003 &  0.00 &     94$\pm$   8 &      9$\pm$   1 &    430$\pm$   4 &     91$\pm$   1 &     93$\pm$   4 &     19$\pm$   1 &     34$\pm$   3 &    404$\pm$  24 &     81$\pm$   7 &   4 \\
J002010.68-004852.8  & 51900  &  390  & 226  &   0.14129$\pm$  0.00002 & 16.07 &    182$\pm$   7 &   -118$\pm$   8 &    127$\pm$   8 &     94$\pm$   6 &     92$\pm$   6 &      4$\pm$   1 &      5$\pm$   1 &     73$\pm$  14 &     69$\pm$  12 &   4 \\
\cutinhead{unknown-type}
J000218.20-110139.5  & 52143  &  650  &  81  &   0.11959$\pm$  0.00002 & 16.76 &    172$\pm$   7 &   -119$\pm$  10 &    125$\pm$  15 &     80$\pm$   8 &    114$\pm$  12 &     13$\pm$   3 &     16$\pm$   3 &     73$\pm$  15 &     87$\pm$  14 &   5 \\
J000503.84+010629.6  & 52559  &  669  & 364  &   0.23857$\pm$  0.00007 & 18.79 &    152$\pm$  21 &    -86$\pm$  19 &    199$\pm$  10 &    123$\pm$  15 &    104$\pm$   7 &      3$\pm$   1 &      3$\pm$   1 &     43$\pm$   9 &     65$\pm$   9 &   5 \\
J001552.74-085430.9  & 52138  &  652  & 409  &   0.13692$\pm$  0.00008 & 17.39 &    282$\pm$  22 &   -181$\pm$  18 &    239$\pm$   9 &    217$\pm$  16 &    117$\pm$   9 &      5$\pm$   1 &     13$\pm$   3 &    132$\pm$  26 &     76$\pm$  16 &   5 \\
J001746.35-005052.7  & 52518  &  687  &  45  &   0.19020$\pm$  0.00004 & 17.82 &    170$\pm$  11 &   -124$\pm$  15 &    151$\pm$  17 &     88$\pm$  13 &    103$\pm$  17 &      2$\pm$   1 &      3$\pm$   1 &     29$\pm$  10 &     28$\pm$   9 &   5 \\
\enddata
\tablecomments{Table 4 is available in its entirety via the link to the machine-readable version above.\\
See note in Table 3.}
\end{deluxetable}

\begin{deluxetable}{llrrrllccl}
\tablecolumns{10}
\tablewidth{0pc}
\tablecaption{List of $pp$-AGN (110) and asym-NEL AGN (235) with strong \oiii wings}
\tabletypesize{\tiny}
\tablehead{
\colhead{No.}&\colhead{SDSS name} &  \colhead{mjd} & \colhead{plate} & \colhead{fiber} & \colhead{$z$} &
\colhead{$r$} & \colhead{$\Delta V_{\rm [OIII]wing}$} & \colhead{$\log L_{\rm [OIII]wing}$} & \colhead{type} \\
\colhead{(1)} & \colhead{(2)}   & \colhead{(3)}    & \colhead{(4)}   & \colhead{(5)}   & \colhead{(6)} &
\colhead{(7)} & \colhead{(8)}    & \colhead{(9)}   & \colhead{(10)} }
\startdata
  1 & J011935.64$-$102613.2 & 52163 &  661 & 285 &  0.124940 & 17.32 &   80.3 & 41.69 & type I $pp$-AGN         \\
  2 & J040001.59$-$065254.1 & 51908 &  464 & 104 &  0.170160 & 16.87 &  132.2 & 41.96 & type I $pp$-AGN         \\
  3 & J080329.02$+$532233.5 & 53384 & 1871 & 290 &  0.219440 & 17.79 &  181.4 & 41.35 & type I $pp$-AGN         \\
  4 & J081203.34$+$571356.7 & 53386 & 1872 & 416 &  0.087510 & 15.67 & -112.5 & 41.61 & type I $pp$-AGN         \\
  5 & J082256.27$+$332201.4 & 52325 &  862 & 495 &  0.138940 & 17.28 &  -67.2 & 41.28 & type I $pp$-AGN         \\
  6 & J084634.20$+$312944.8 & 52991 & 1270 & 438 &  0.120960 & 16.80 & -138.5 & 40.84 & type I $pp$-AGN         \\
  7 & J085944.13$+$295408.4 & 53357 & 1934 & 365 &  0.048770 & 15.43 &  -82.3 & 39.91 & type I $pp$-AGN         \\
  8 & J091544.19$+$300922.0 & 53379 & 1938 & 309 &  0.129380 & 16.47 &  259.2 & 41.88 & type I $pp$-AGN         \\
  9 & J093549.73$+$072502.2 & 52733 & 1196 & 106 &  0.213880 & 17.14 &  -86.8 & 41.18 & type I $pp$-AGN         \\
 10 & J095950.32$-$001339.0 & 51633 &  268 & 146 &  0.090670 & 16.83 &  -41.0 & 41.31 & type I $pp$-AGN         \\
...  & ...                      &  ...       & ...    & ...     & ...    & ...                \\
\enddata
\tablecomments{Table 5 is available in its entirety via the link to the machine-readable version above.\\
Columns (1)-(7) are No., SDSS name, mjd, plate, fiber, redshifts and $r$-band magnitude, respectively.
Column (8) is shift of the strong wings in \kms, and (9) is the luminosity in \ergs. Column (10) indicates
the types of the objects. }
\end{deluxetable}
\begin{deluxetable}{lllcrrcl}
\tablecolumns{8}
\tablewidth{0pc}
\tablecaption{List of 54 dual-cored \ppgalaxies\ and asym-NEL galaxies}
\tabletypesize{\tiny}
\tablehead{
\colhead{No.}&\colhead{SDSS name} & \colhead{mjd} & \colhead{plate} & \colhead{fiber} &
\colhead{$z$} & \colhead{$r$} &\colhead{type}}
\startdata
   1  & J030558.72$-$005027.5 & 52289 &  802 &  71 &  0.154710 & 19.13 & 2-type II  \\
   2  & J075055.44$+$354411.4 & 52017 &  543 & 224 &  0.135580 & 17.10 & unknown     \\
   3  & J103618.74$+$152310.1 & 54177 & 2594 & 244 &  0.065870 & 15.80 & unknown     \\
   4  & J103801.87$+$371643.2 & 53432 & 1973 & 625 &  0.119080 & 16.22 & 2-type II  \\
   5  & J111834.31$+$294110.7 & 53793 & 2215 & 543 &  0.138940 & 18.11 & type II+SF \\
   6  & J112555.93$+$142206.4 & 53385 & 1754 & 317 &  0.208830 & 18.78 & type II+SF \\
   7  & J121900.51$+$152442.2 & 53436 & 1767 & 344 &  0.241500 & 17.72 & 2-SF       \\
   8  & J122222.25$+$022038.6 & 52282 &  518 &  28 &  0.111020 & 16.40 & unknown     \\
   9  & J122239.58$+$134638.5 & 53436 & 1767 & 218 &  0.080020 & 16.71 & unknown     \\
  10  & J130242.81$+$143405.4 & 53498 & 1771 & 111 &  0.142160 & 16.91 & unknown    \\
  11  & J135224.16$+$181144.3 & 54508 & 2756 & 262 &  0.168460 & 16.71 & unknown    \\
  12  & J143615.95$+$560124.1 & 52668 & 1162 & 448 &  0.140700 & 17.49 & unknown    \\
  13  & J144420.56$+$120742.9 & 53531 & 1712 & 158 &  0.029140 & 15.44 & unknown    \\
  14  & J145059.71$-$000215.2 & 51994 &  309 & 304 &  0.043080 & 13.85 & unknown    \\
  15  & J164648.33$+$262857.4 & 53260 & 1691 & 311 &  0.137350 & 16.76 & 2-type II  \\
  16  & J002856.80$+$004037.7 & 51782 &  391 & 534 &  0.090060 & 16.68 & type II+SF \\
  17  & J021242.27$+$002903.8 & 53763 & 1507 & 421 &  0.150020 & 17.48 & type II+SF \\
  18  & J034508.79$-$005546.6 & 54465 & 2639 &  44 &  0.226760 & 18.22 & unknown    \\
  19  & J080030.74$+$440720.4 & 51876 &  437 & 361 &  0.131870 & 16.60 & 2-SF       \\
  20  & J081948.05$+$254329.0 & 52962 & 1585 & 450 &  0.082080 & 18.08 & type II+SF \\
  21  & J094410.57$+$654744.8 & 54468 & 1788 & 442 &  0.138870 & 17.31 & type II+SF \\
  22  & J100443.64$+$062555.0 & 52731 &  995 & 600 &  0.163760 & 19.17 & 2-type II  \\
  23  & J102700.40$+$174901.0 & 54140 & 2591 & 480 &  0.066580 & 15.20 & unknown     \\
  24  & J102938.95$+$375030.0 & 53432 & 1973 & 405 &  0.151760 & 19.64 & 2-type II  \\
  25  & J103846.97$+$155359.0 & 54177 & 2594 & 498 &  0.194230 & 17.82 & type II+SF \\
  26  & J105045.75$+$664146.9 & 51929 &  490 & 430 &  0.098970 & 16.43 & unknown    \\
  27  & J111506.62$+$132913.0 & 53379 & 1752 &  54 &  0.169970 & 16.55 & unknown    \\
  28  & J112756.91$+$245934.1 & 54154 & 2497 & 593 &  0.115100 & 17.03 & 2-type II  \\
  29  & J112850.25$+$190339.8 & 54180 & 2502 & 314 &  0.166330 & 17.82 & 2-SF       \\
  30  & J114617.47$+$145305.2 & 53415 & 1762 & 263 &  0.173420 & 18.05 & 2-type II  \\
  31  & J123656.65$+$254129.8 & 54498 & 2659 & 611 &  0.087220 & 16.06 & 2-type II  \\
  32  & J124132.12$+$535257.3 & 52673 & 1038 & 298 &  0.083270 & 15.80 & 2-type II  \\
  33  & J135046.61$+$315138.4 & 53503 & 2024 &  20 &  0.105600 & 17.33 & unknown    \\
  34  & J135313.15$+$475256.9 & 52736 & 1284 &  38 &  0.124390 & 16.61 & unknown    \\
  35  & J135830.04$+$182736.1 & 54509 & 2757 & 382 &  0.063010 & 15.88 & type II+SF \\
  36  & J140247.81$+$144900.8 & 54272 & 2744 & 124 &  0.225760 & 16.74 & unknown    \\
  37  & J142032.07$+$020351.5 & 51994 &  533 &  35 &  0.068740 & 15.87 & 2-type II  \\
  38  & J142606.65$+$202831.6 & 54552 & 2787 & 183 &  0.076810 & 16.12 & type II+SF \\
  39  & J143246.31$-$020617.7 & 52409 &  919 & 275 &  0.056270 & 15.11 & 2-type II  \\
  40  & J143520.64$+$185842.7 & 54535 & 2775 & 431 &  0.180770 & 17.41 & 2-type II  \\
  41  & J145320.90$+$060626.0 & 54560 & 1830 & 426 &  0.093900 & 16.75 & type II+SF \\
  42  & J145857.24$+$090232.4 & 53884 & 1815 & 377 &  0.137920 & 16.86 & 2-SF       \\
  43  & J151109.24$-$015229.7 & 52404 &  923 & 238 &  0.124740 & 17.67 & unknown    \\
  44  & J151706.75$+$060645.7 & 54561 & 1833 & 545 &  0.044910 & 15.83 & 2-SF       \\
  45  & J151713.62$+$331008.6 & 53116 & 1386 & 173 &  0.082200 & 16.05 & 2-type II  \\
  46  & J154412.23$+$274955.6 & 53534 & 1653 & 154 &  0.116290 & 17.72 & 2-SF       \\
  47  & J155708.82$+$273518.7 & 52822 & 1392 & 400 &  0.124170 & 16.27 & unknown    \\
  48  & J160102.81$+$070612.1 & 53858 & 1729 & 400 &  0.139390 & 17.61 & type II+SF \\
  49  & J161037.33$+$073852.3 & 53498 & 1730 & 180 &  0.130230 & 17.43 & type II+SF \\
  50  & J204631.42$+$001708.8 & 52466 &  982 & 416 &  0.137170 & 17.25 & 2-SF       \\
  51  & J205708.39$+$010805.2 & 52442 &  984 & 369 &  0.060500 & 16.44 & unknown    \\
  52  & J220329.87$+$123327.6 & 52519 &  735 & 302 &  0.151040 & 17.62 & 2-type II  \\
  53  & J220804.49$+$010805.8 & 51788 &  373 & 523 &  0.085490 & 15.56 & unknown    \\
  54  & J221924.98$-$093821.6 & 52203 &  719 &   7 &  0.094870 & 15.95 & 2-SF
\enddata
\tablecomments{The images and the spectra of the 54 galaxies are shown in Figure 11.
The number sequence of the images should be from {\it left} to {\it right} and from {\it top} to
{\it bottom}. No. 1-15 are \ppgalaxies\ and 16-54 are asym-NEL galaxies.}
\end{deluxetable}
\begin{deluxetable}{rlcrrccclcrrccc}
\rotate
\tablecolumns{15}
\tablewidth{0pc}
\tablecaption{List of 255 galaxies with dual cores separated $\ge 3^{\prime\prime}$}
\tabletypesize{\tiny}
\tablehead{
  &  \multicolumn{6}{c}{primary core} & & \multicolumn{6}{c}{secondary core} \\\cline{2-7} \cline{9-14}
\colhead{No.}&\colhead{SDSS name} & \colhead{mjd} & \colhead{plate} & \colhead{fiber} & \colhead{$z$} & \colhead{$r$} & \colhead{} &
\colhead{SDSS name} & \colhead{mjd} & \colhead{plate} & \colhead{fiber} & \colhead{$z$} & \colhead{$r$} & \colhead{Note}}
\startdata
  1 & J090012.70+183439.2 & 53729 & 2283 & 160 & 0.08118 & 16.97 & & J090012.30+183436.8 & 53700 & 2285 & 316 & 0.08040 & 17.17   & normal galaxy       \\
  2 & J101455.13+003337.5 & 51909 &  270 & 586 & 0.18723 & 17.16 & &                     &       &      &     &         &       &                   \\
  3 & J135432.27+582345.8 & 52668 & 1158 & 517 & 0.06061 & 14.34 & &                     &       &      &     &         &       &                   \\
  4 & J000249.07+004504.8 & 51793 &  388 & 345 & 0.08663 & 16.08 & & J000249.44+004506.7 & 52203 &  685 & 593 & 0.08653 & 16.43   & normal galaxy       \\
  5 & J081423.00+033628.3 & 52641 & 1184 & 230 & 0.17077 & 19.28 & &                     &       &      &     &         &       &                   \\
  6 & J081705.22+032146.2 & 52641 & 1184 &  53 & 0.09329 & 17.11 & &                     &       &      &     &         &       &                   \\
  7 & J091954.54+325559.8 & 52990 & 1592 & 160 & 0.04928 & 17.43 & &                     &       &      &     &         &       &                   \\
  8 & J095900.10+293337.6 & 53436 & 1950 & 218 & 0.17887 & 17.54 & &                     &       &      &     &         &       &                   \\
  9 & J100602.14+071131.0 & 52751 & 1236 &  54 & 0.12148 & 18.62 & & J100602.51+071131.8 & 52641 &  996 & 328 & 0.12051 & 15.77   & normal galaxy       \\
 10 & J102207.37-000415.5 & 51883 &  271 &  39 & 0.12547 & 17.26 & &                     &       &      &     &         &       &                   \\
\enddata
\tablecomments{Table 7 is available in its entirety via the link to the machine-readable version above.\\
$-$ It should be noted that SDSS fibers only cover $3^{\prime\prime}$ and the double-peaked profiles of
these source {\it only} originate from {\it one} of the cores marked by red square in the images given
by Figure 12 as shown by Appendix C.\\
$-$ There are 39 sources with SDSS spectra observation of the secondary core, and have $\Delta z<0.0017$
(500 km/s). The last column shows the property of the secondary core: normal (single-peaked) galaxy,
$pp$-galaxy, asym-NEL galaxy. }
\end{deluxetable}

\clearpage

\begin{figure}
{\centering
\figurenum{1}
\includegraphics[angle=0.0,scale=0.75]{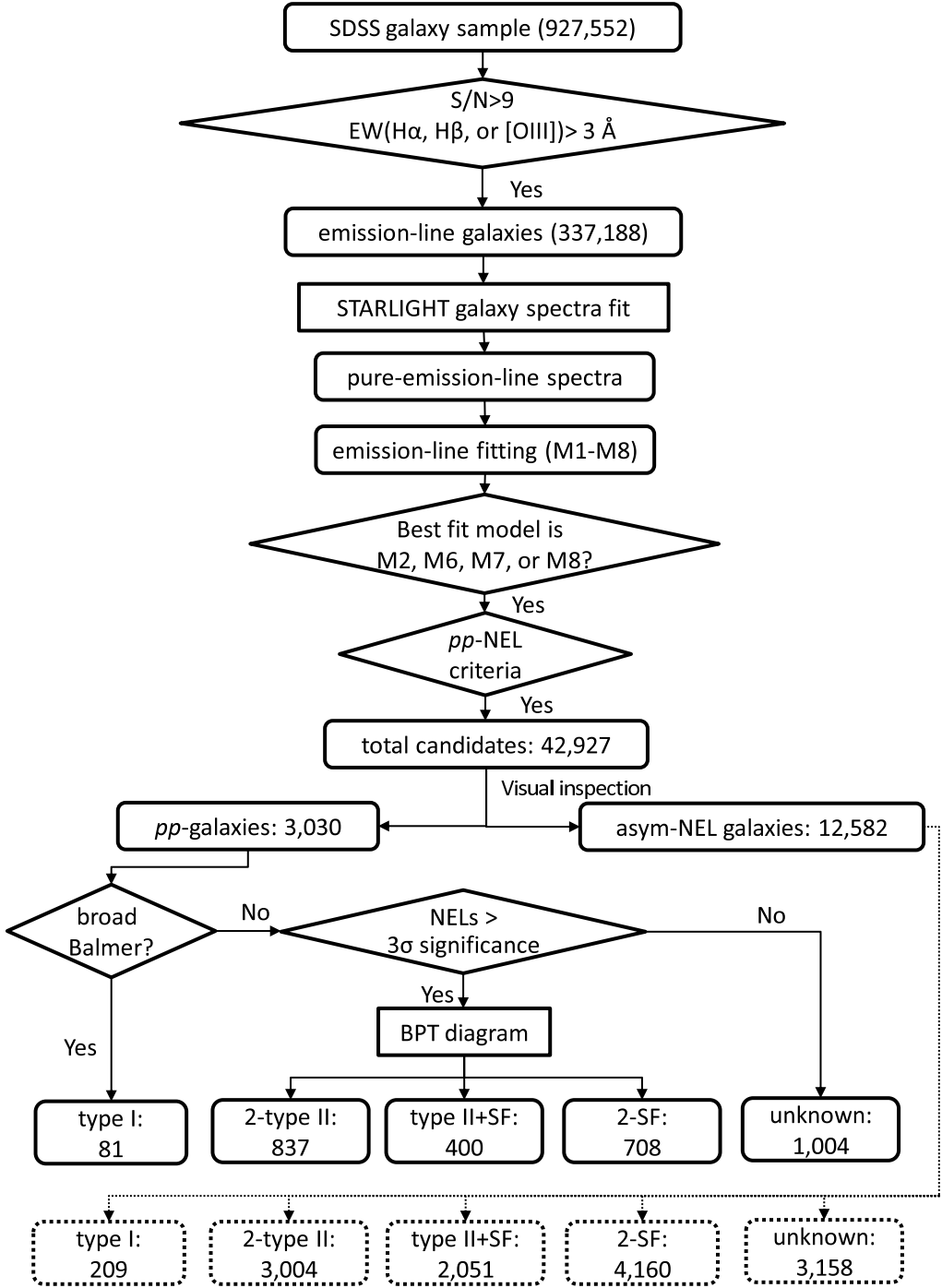}
\figcaption{\footnotesize Flowchart of sample selection. M1-M8 are the eight models used for fitting the
pure emission-line spectra. The dashed line shows that asym-NEL galaxies are also classified by the same
method as double-peaked sources. The numbers indicated in squares are the amount of the corresponding
objects.}
\label{fig1}
}
\end{figure}

\clearpage

{\centering
\figurenum{2}
\includegraphics[angle=-90.0,scale=0.65]{fig2.ps}
\figcaption{\footnotesize
Example of galactic spectra (SDSS J000656.85+154847.9) fitting with stellar templates
(V04, {\it top left}) by using the direct pixel-fitting method and with stellar population
templates (BC03, {\it bottom left}) by using the STARLIGHT code. The green and blue lines
are the observed spectra. During the absorption line fitting, the blue lines are masked
to eliminate the influence of the emission lines and we only use the points labeled as green
for fitting. The red lines are the fitted stellar spectra. The black lines represent the
residual spectra. The corresponding $\chi^2$ distributions of the fits with the direct pixel-fitting
method ({\it top right}) and the STARLIGHT code ({\it bottom right}) are also shown. The black
lines represent the $\chi^2$ calculated from the total fitting range. The red lines represent
the $\chi^2$ calculated in the restricted region 3910-4060\AA\ and 5250-5820\AA\ of the total
fitting range, which covers the Ca {\sc ii} H+K and Mg {\sc i}b triplet absorption lines.
}
\label{cd_fitting}
}

\clearpage

{\centering
\figurenum{3}
\includegraphics[angle=0.0,scale=0.8]{fig3.ps}
\figcaption{\footnotesize Example of a pure emission-line spectrum after subtracting the stellar
contribution of its host galaxy. We fit the spectrum with different emission
line components (SDSS J123605.45-014119.2). The black lines represent the pure emission-line spectrum,
the blue lines show one narrow Gaussian group, and the red lines show another narrow Gaussian group,
the pink line stands for the broader \oiii wings, while the brown lines represent the broad Balmer lines
(H$\alpha$ and H$\beta$). The green lines shows the final fitting model.}
\label{fig3}
}
%

{
\centering
\figurenum{4}
\includegraphics[angle=-90.0,scale=0.68]{fig4.ps}
\figcaption{\footnotesize BPT diagrams of blue and red components of $pp$-galaxies $(a_1, b_1)$ and asym-NEL
galaxies $(a_2, b_2)$. The solid and dashed lines correspond to the dividing lines
between AGN and SF galaxies given by Kewley et al. (2001) and Kauffmann et al. (2003), respectively. The
blue cross symbols in panel $(a_1)$ and $(a_2)$ correspond to the blue component and the red rectangulars
represent the red component of NELs of type I AGN. The blue and red points in panel $(b_1)$ and $(b_2)$
correspond to the blue and red components of the NELs of narrow-line galaxies respectively.}
\label{fig4}
%
}

\clearpage

\begin{figure}
\centering
\figurenum{5}
\includegraphics[angle=0.0,scale=1.0]{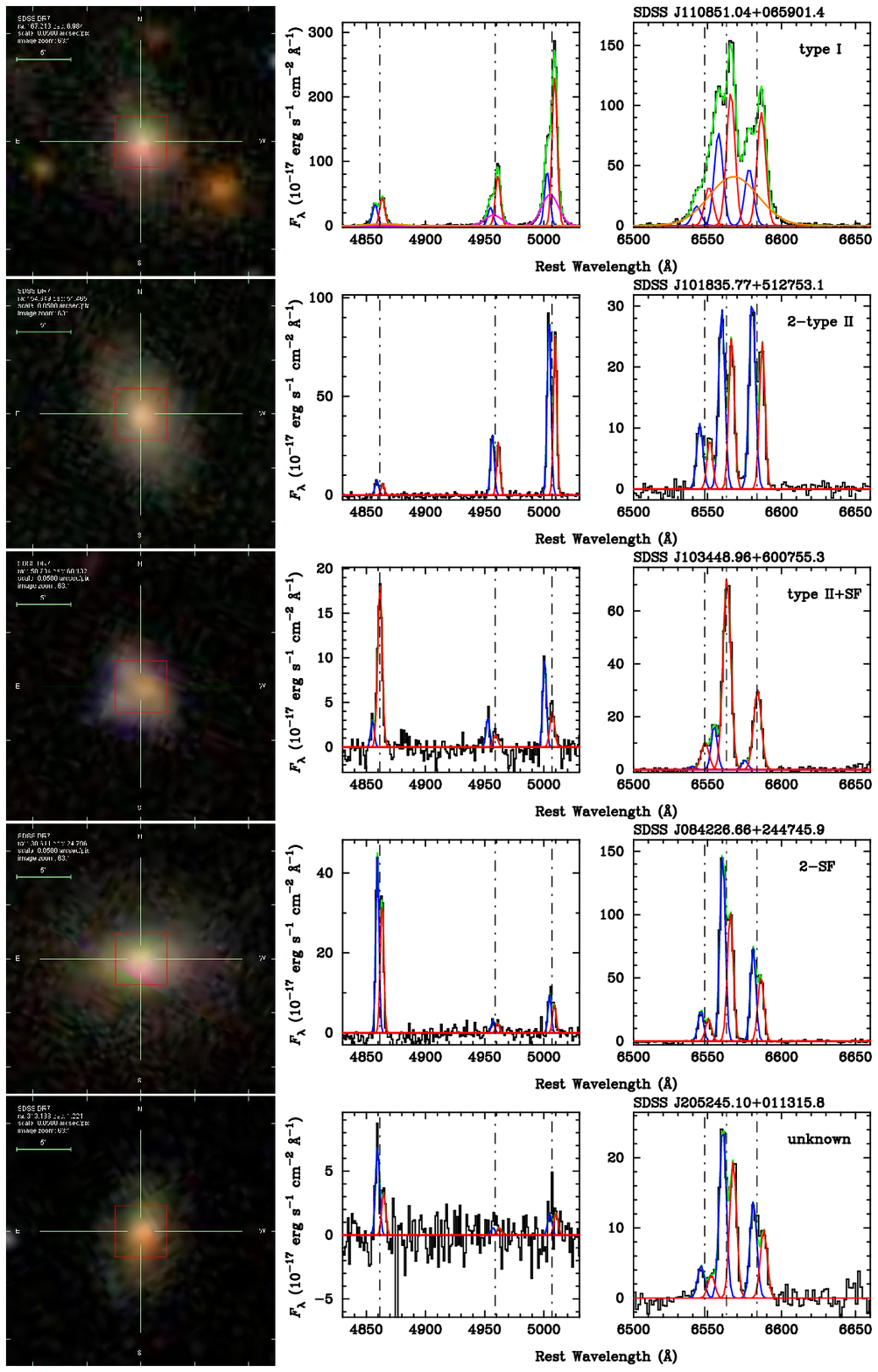}
\figcaption{\footnotesize Image and spectrum examples of five kinds of {\it pp}-galaxies, showing
the complicated double-peaked components.}
\label{fig5}
\end{figure}

\begin{figure}
\centering
\figurenum{6}
\includegraphics[angle=0.0,scale=1.0]{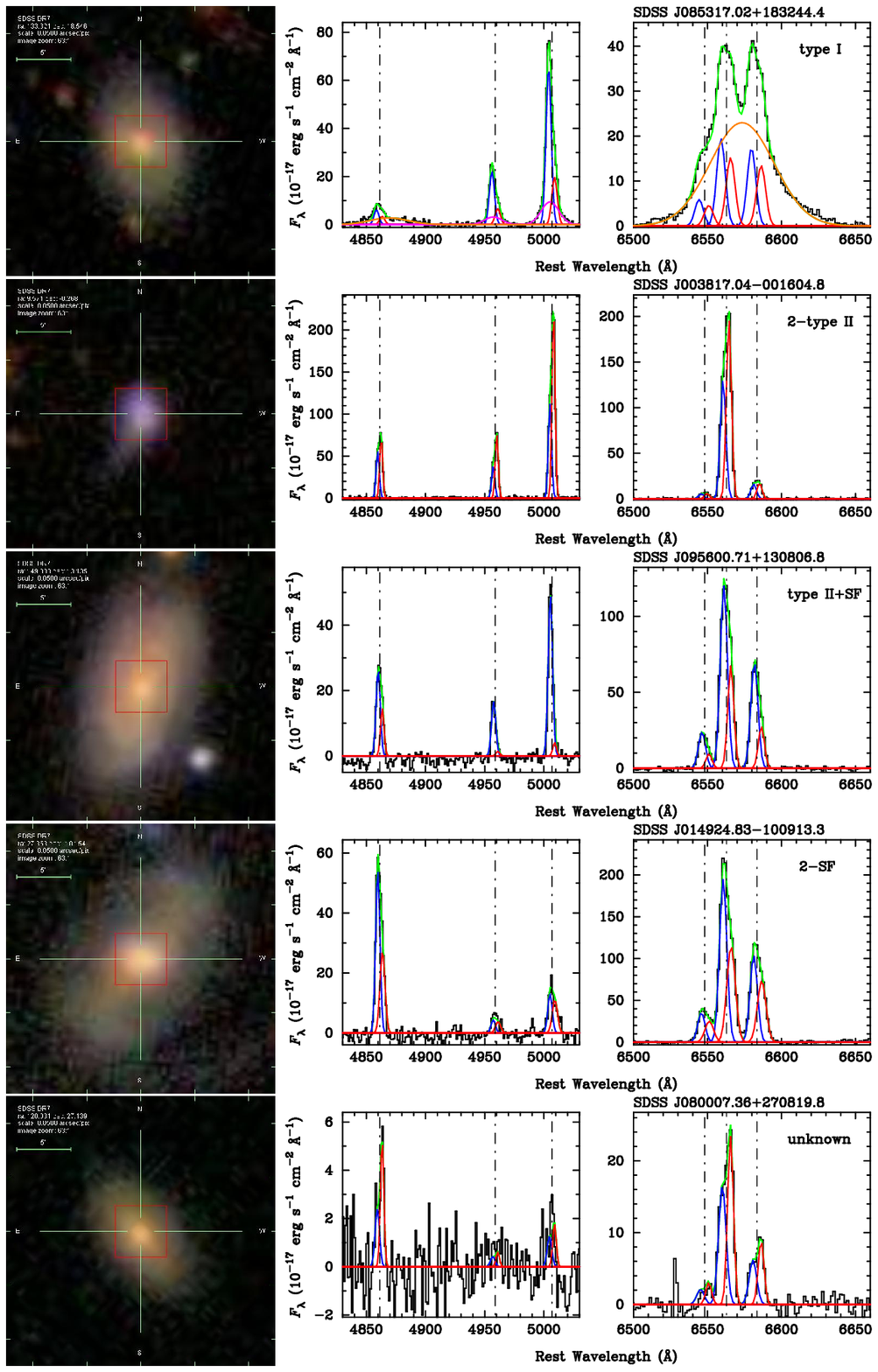}
\caption{\footnotesize Image and spectrum examples of asym-NEL galaxies. They are candidates
of {\it pp}-galaxies.}
\label{fig6}
\end{figure}

\clearpage
{
\centering
\figurenum{7}
\includegraphics[angle=-90.0,scale=0.65]{fig7.ps}
\figcaption{\footnotesize ({\it a}): The redshift distributions of the different kinds of {\it pp}-galaxies
and ({\it b}) for asym-NEL galaxies; ({\it c}) for the 54 dual-cored galaxies
(red line, with dual cores separated $< 3^{\prime\prime}$ listed in Table 6) along with the 255 dual-cored
galaxies (the blue line, with dual cores separated $\ge 3^{\prime\prime}$ listed in Table 7). }
\label{redshifts}
}


{
\centering
\figurenum{8}
\includegraphics[angle=-90.0,scale=0.6]{fig8.ps}
\figcaption{\footnotesize The shift distributions of the red and blue peaks of $pp$-galaxies ($top$)
and asym-NEL galaxies ($bottom$). The blue line in each panel corresponds to the distribution of $\dvb$ and
the red line represent that of $\dvr$. In each panel, the blue dashed line represents the averaged blue shifts
($\langle \dvb\rangle$) whereas the red dashed line stands for averaged red shifts $(\langle \dvr\rangle)$.
The mean values (dashed line) are equal to the median values (dotted line) in these distributions.
$\langle {\rm log} ~v \rangle= \langle {\rm log} ~|\dvr / \dvb| \rangle$ is the averaged logarithmic shift
ratio of the red and blue components. Note that $\langle \dvr \rangle$ and $\langle \dvb \rangle$ given inside
the panels are in units of $10^2$\kms.}
\label{fig8}
}

\clearpage
{
\centering
\figurenum{9{\it a}}
\includegraphics[angle=-90.0,scale=0.6]{fig9a.ps}
\figcaption{\footnotesize The $\log\left(F_{\rm H\alpha}^{\rm r}/F_{\rm H\alpha}^{\rm b}\right)$ distributions.
The upper panel is for \ppgalaxies\ and lower panel for asym-NEL galaxies. The averaged values
of the ratios are indicated in every plots. Here
$\langle \log f \rangle = \langle \log (F^r_{\rm H\alpha} /F_{\rm H\alpha}^b) \rangle$.}
\figurenum{9{\it b}}
\includegraphics[angle=-90.0,scale=0.6]{fig9b.ps}
\figcaption{\footnotesize The $\log\left(F_{\rm [O~III]}^{\rm r}/F_{\rm [O~III]}^{\rm b}\right)$ distributions.
The upper panel is for \ppgalaxies\ and lower panel for asym-NEL galaxies. Comparing the distributions,
we find that distributions of \oiii\ flux ratios are generally broader than H$\alpha$. Here
$\langle {\rm log} f \rangle = \langle {\rm log} (F^r_{\rm [O~III]} /F_{\rm [O~III]}^b) \rangle$.}
\label{fig9}
}

\clearpage

\appendix

\section{Monte-Carlo simulation of the selection criteria}
Monte Carlo simulations
have been performed in order to define the final selection criteria that allow us to maximally select
potential galaxies with double-peaked narrow emission lines. According to the SDSS spectra resolution,
we generate artificial spectra with a wavelength interval of 1\AA\ in the wavelength range of [4950,5050]\AA\,
which only contains the \oiii$\lambda$5007 narrow emission line. In order to show how the criteria
depend on the spectral noise, we do the following two steps.

First, two Gaussian components are generated randomly to compose the \oiii$\lambda$5007 emission line.
Table 8 gives the simulated ranges of the parameters of the \oiii\ lines. For a random combination of the two
components, double-peaked profiles do not necessarily appear in the composite spectra. However, it
is easier to judge the presence of double-peaked profiles for a simulated noise-free spectrum. We search
a critical wavelength point ($\lambda_i$) of the spectrum, where $F(\lambda_i)$ is smaller than
its several neighbors at both sides, namely, $F(\lambda_{i-j})>F({\lambda_i})<F(\lambda_{i+j})$, where
$j=1,2,3$. With the presence of $\lambda_i$, the spectrum is classified as an intrinsic
double-peaked profile. Otherwise, it is a single-peaked spectrum. We generate many composed
spectra ($\sim 1000$) which are intrinsically double-peaked and noise-free.

\begin{deluxetable}{lll}
\tablecolumns{3}
\tablewidth{0pc}
\tablecaption{Monte-Carlo simulations: ranges of double-peaked \oiii\ lines}
\tabletypesize{\footnotesize}
\tablehead{
\colhead{Parameters} & \colhead{ranges} & \colhead{notes}}
\startdata
$F_{\rm b}/F_{\rm r}$    & [$10^{-2},10^2$]   & constrained by SDSS signal-noise ratio \\
$V_{\rm b}$              & [$-300,0$]\kms       & determined by the observed range       \\
$V_{\rm r}$              & [$0,300$]\kms        &                                        \\
FWHM$_{\rm r,b}$         & [$150,700$]\kms      & determined by SDSS spectral resolution \\
S/N                      & 10 or 30           & by SDSS signal-to-noise ratio
\enddata
\end{deluxetable}

Second, we examine if the produced spectra are double-peaked after adding noise to them. We generate
two kinds of noisy spectra with ${\rm S/N}=10$ and ${\rm S/N}=30$. Through visual inspection,
we select noisy spectra which show double-peaked profiles. We then fit these double-peaked spectra
with the same methods as described in Section 2.2. We find that $\sim 800$ and $\sim 900$ can be selected
as double-peaked objects from the 1,000 simulated-spectra. This means only 80\% and 90\% of the intrinsically
double-peaked objects can be found for ${\rm S/N}=10$ and ${\rm S/N}=30$, respectively. The
distributions of the parameters of double-peaked spectra yield the selection criteria.

We use three quantities $(\Delta V /{\rm FWHM}_{\rm min}$, $F_{\rm r}/F_{\rm b})$
and S/N in the simulations, where
$\Delta V=\Delta V_{\rm r}-\Delta V_{\rm b}$ is the velocity separation of the red and blue components,
and ${\rm FWHM_{min}}$ is the minimum of ${\rm (FWHM_{red}, FWHM_{blue})}$. For a fixed S/N, we plot all the
visually-selected double-peaked spectra in $F_{\rm r}/F_{\rm b}-\Delta V /{\rm FWHM}_{\rm min}$ plane.
Then the reasonable ranges of $F_{\rm r}/F_{\rm b}$ and $\Delta V /{\rm FWHM}_{\rm min}$ can be easily
determined by the plane. Figure 10 {\it left} shows the simulations for two cases with
${\rm S/N}=10$ and ${\rm S/N}=30$, respectively.

From Figure 10 {\it left} panel, we find that flux ratios beyond the range of $[0.05, 20]$ can not
be detected for ${\rm S/N}=30$. Therefore, we conservatively use flux ratios of $[0.03, 30]$ in practice.
Most double-peaked objects have $\Delta V /{\rm FWHM}_{\rm min}>1$. To make a proper
complete automatic search, we use a criterion as $\Delta V /{\rm FWHM}_{\rm min}>0.8$.

{
\centering
\figurenum{10}
\includegraphics[angle=-90.0,scale=0.6]{fig10.ps}\\%
\figcaption{\footnotesize Monte Carlo simulation for the selection criteria of double-peaked,
asymmetric and top-flat narrow emission lines with the same value of the parameters given by
Table 8. The classifications of the simulated profiles are visually inspected.
The black points represent the spectra with ${\rm S/N}=10$ and the red points
correspond to ${\rm S/N}=30$. The vertical two lines represent $\Delta V /{\rm FWHM}_{\rm min}=0.8$
(dashed line) and 1.0 (dotted line). The {\it left} panel shows simulations for apparent
double-peaked profiles of emission lines whereas the {\it right} panel for asym-NEL galaxies.}
\label{mcsimulation}
}

During the visual inspection of the simulated spectra, we find many asym-NEL sources which are
referred to those sources without ``trough". We plot them in the {\it right} panel of Figure 10.
The Monte-Carlo simulations for asym-NEL sources show that: 1) they have a similar range of
$F_r/F_b$ with double-peaked sources; 2) but they have different ranges of $\Delta V/{\rm FWHM_{min}}$.
Therefore, the selected sources unavoidably cover many asym-NEL sources.

The present criteria guarantee the completeness of double-peaked sources. We emphasis that the criteria
are only fit to selection of double-peaked sources rather than to the asym-NELs, but asym-NEL sources
are selected as by-product of the criteria. For a
goal of selecting a complete sample of asym-NEL galaxies, we have to define additional criteria for them,
such as a definition of profile asymmetry, and try to find reasonable criteria. We will do this in a
future paper. On the other hand, realistic distributions of parameters controlling the profiles listed
in Table 8 are unknown, appearance of different kinds of profiles can not be predicted. It is thus
hard for the Monte-Carlo simulations to predict the completeness of asym-NEL sources.

We also simulate the probability of mis-selection that single NELs could be selected as double-peaked
or asym-NELs by our criteria.
For S/N=30, only 2 of 10,000 are selected as asym-NEL sources, none as double-peaked sources whereas
for S/N=10, only 1 as the double-peaked and 2 as the asym-NEL. It is clear that the fraction of mis-selected
sources is very small.

\section{Images and spectra of 54 dual-cored \ppgalaxies\ and asym-NEL galaxies}
The SDSS optical images and spectra of 54 dual-cored galaxies listed in Table 6 are
shown in Figure 11.
\newpage
{
\centering
\figurenum{11}
\includegraphics[angle=0.0,scale=0.88]{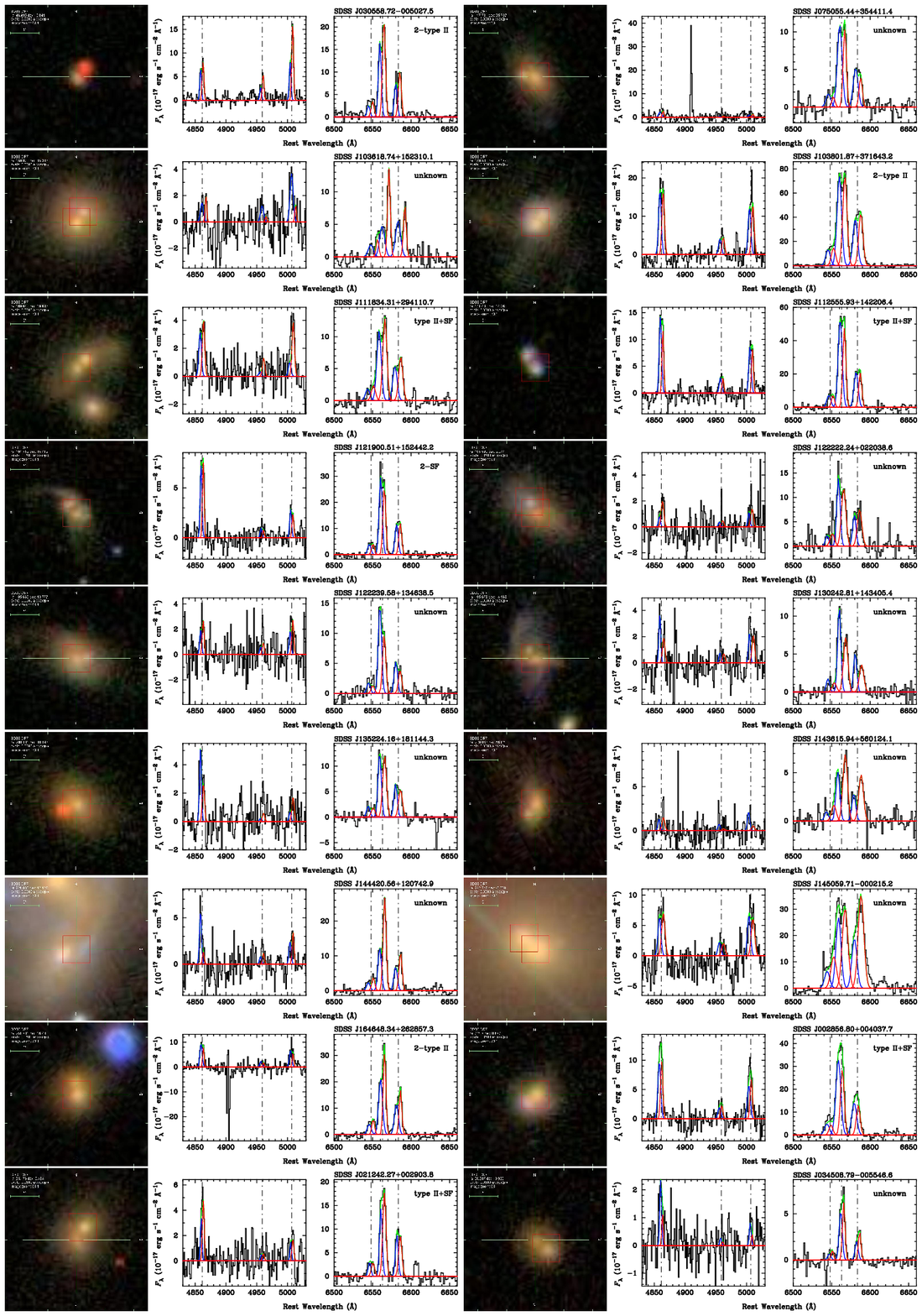}\\%
\figcaption{\footnotesize Images and spectra of 54 $pp$-galaxies and asym-NEL galaxies listed in Table 6.
The name and type of the sources are labeled.  }
\includegraphics[angle=0.0,scale=0.88]{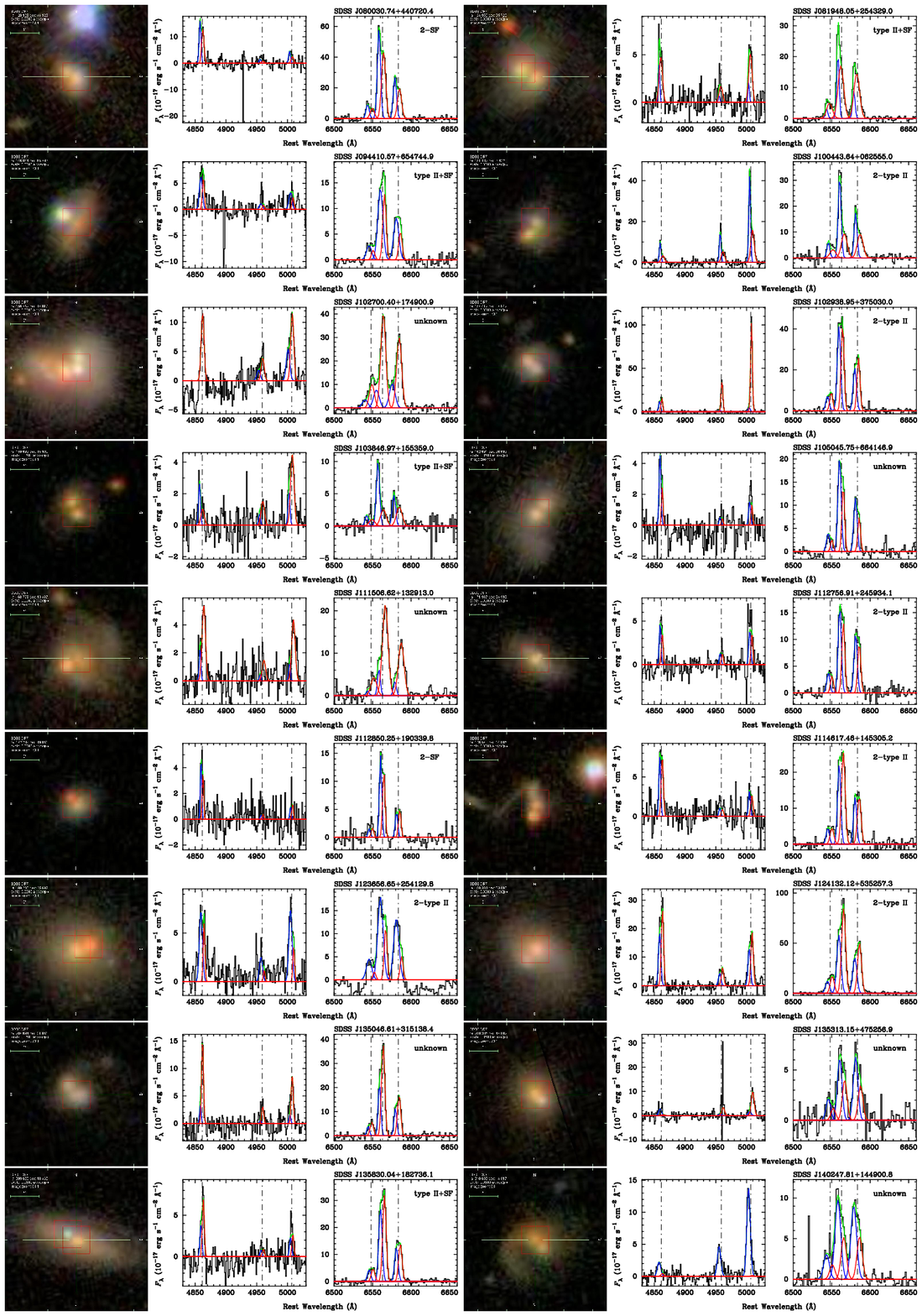}\\%
\figcaption{\footnotesize {\it Continued}}
\includegraphics[angle=0.0,scale=0.88]{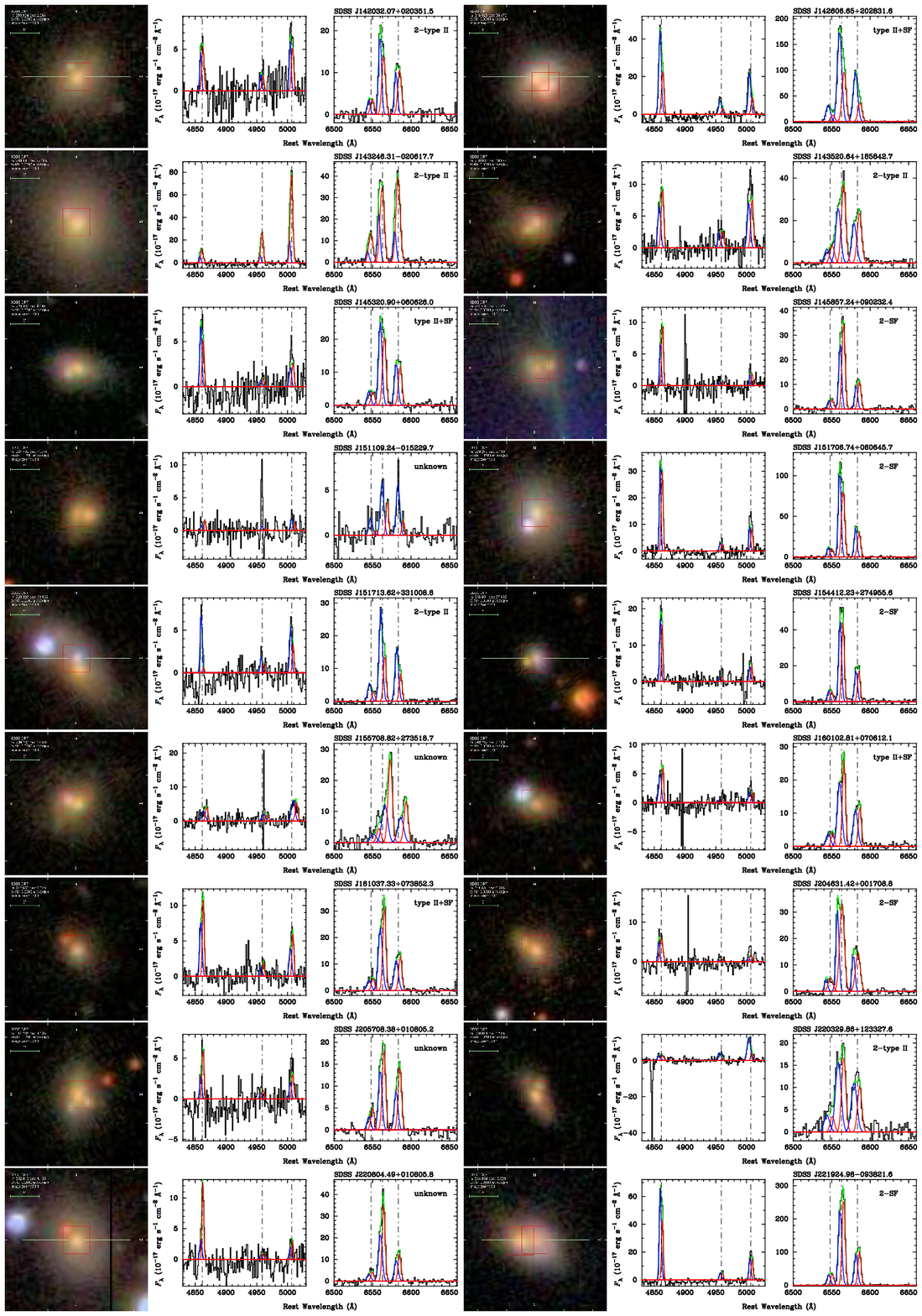}%
\figcaption{\footnotesize {\it Continued}}
}

\newpage
\section{Images of 255 objects with double cores separated by $\ge 3^{\prime\prime}$}
The SDSS optical images of the 255 dual-cored galaxies listed in Table 7 are shown in
Figure 12.

{\centering
\figurenum{12}
\includegraphics[angle=0.0,scale=0.72]{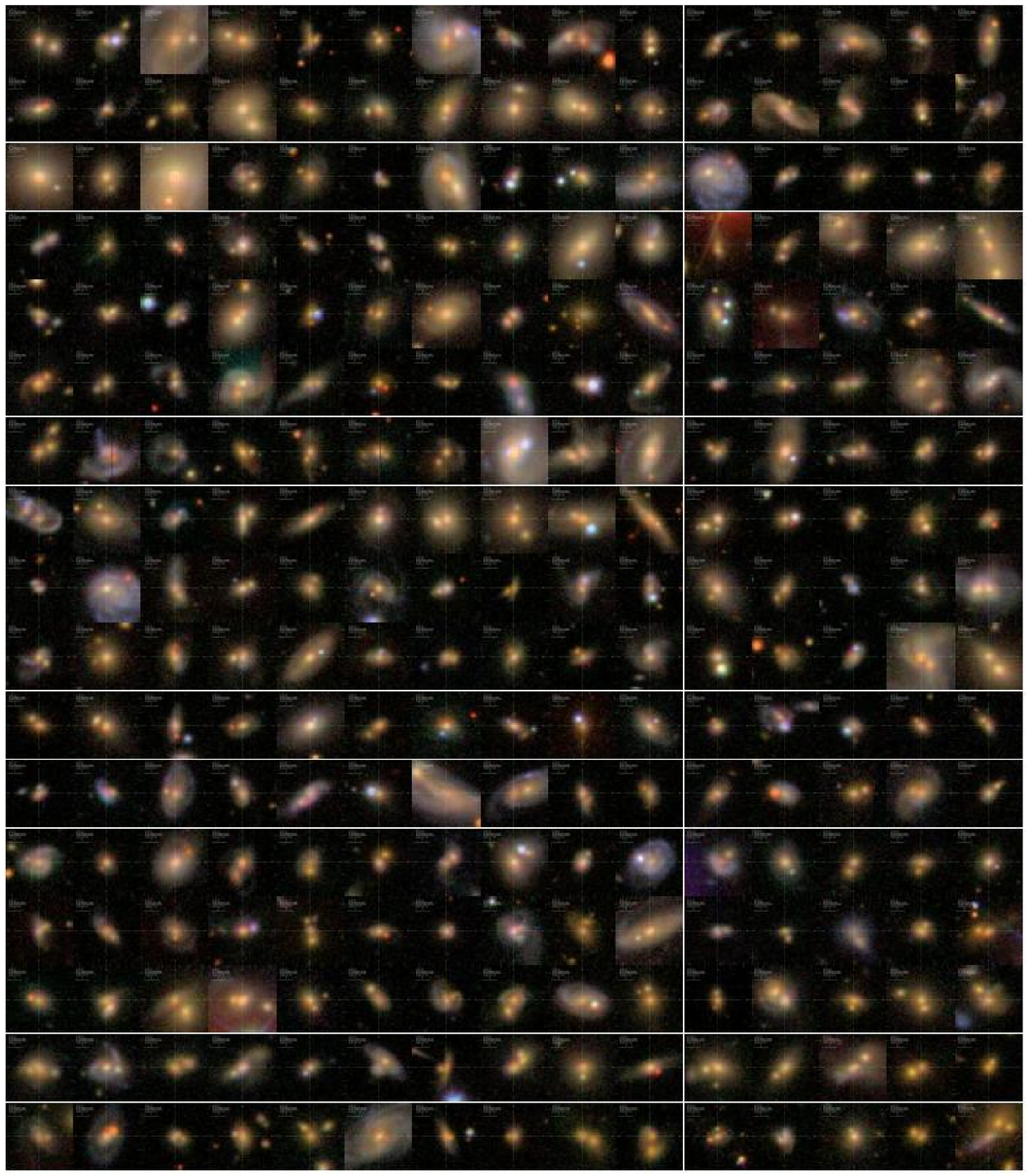}
\figcaption{\footnotesize Image number can be obtained through $n=x+15(y-1)$, where $x=1,2,...15$ is
a row number of an image from {\it left} to {\it right} and $y=1,2,3,...17$ is a line number of an
image from {\it top} to {\it below}. There are 15 images in each row.
The number $n$ corresponds to the number of objects listed in Table 7. No. $n=1-67$ are
those galaxies with double-peaked profiles, and $n=68-255$ are ones with asymmetric or top-flat profiles.
Each image has the same size of $25\arcsec \times 25\arcsec$ and same orientations (i.e. up North,
left East).}
}

\end{document}